\documentclass[12pt]{article}
\usepackage{latexsym}
\usepackage{epsfig,amssymb,euscript,slashed}
\usepackage{amsmath}
\usepackage{cite}
\usepackage{array,calc,epsfig}
\usepackage[linktocpage]{hyperref}
\usepackage{pstricks}
\usepackage{bbm}
\usepackage{fancybox}

\def\be{\begin{equation}}
\def\ee{\end{equation}}
\def\bseq{\begin{subequations}}
\def\eseq{\end{subequations}}

\def\bea{\begin{eqnarray}}
\def\eea{\end{eqnarray}}

\def\bseq{\begin{subequations}}
\def\eseq{\end{subequations}}

\numberwithin{equation}{section} %%
\usepackage[]{graphicx}

\def\d {{\rm d}}
\def\cala         {{\cal A}}

\def\calc         {{\cal C}}

\def\calf         {{\cal F}}
\def\calg         {{\cal G}}

\def\cali         {{\cal I}}
\def\calj         {{\cal J}}
\def\calk         {{\cal K}}
\def\call         {{\cal L}}
\def\calm         {{\cal M}}
\def\caln         {{\cal N}}
\def\calo         {{\cal O}}

\def\calv         {{\cal V}}

\def\del          {\partial}
\def\delbar       {\bar\partial}
\def\ii           {{\rm i}}

\def\Re           {{\rm Re\hskip0.1em}}
\def\Im           {{\rm Im\hskip0.1em}}

\def\sqr#1#2{{\vcenter{\vbox{\hrule height.#2pt
 \hbox{\vrule width.#2pt height#1pt \kern#1pt \vrule width.#2pt}\hrule
 height.#2pt}}}}

\newcommand{\uu}[1]{{\mbox{\tiny $({#1})$}}}

\def\d{\text{d}}

\def\slashchar#1{\setbox0=\hbox{$#1$}           % set a box for #1
\dimen0=\wd0                                 % and get its size
\setbox1=\hbox{/} \dimen1=\wd1               % get siste of /
\ifdim\dimen0>\dimen1                        % #1 is bigger
\rlap{\hbox to \dimen0{\hfil/\hfil}}      % so center / in box
#1                                        % and print #1
\else                                        % / is bigger
\rlap{\hbox to \dimen1{\hfil$#1$\hfil}}   % so center #1
/                                         % and print /
\fi}

\begin{document}
\font\cmss=cmss10 \font\cmsss=cmss10 at 7pt

\begin{flushright}{\scriptsize DFPD-2014/TH/18}
\end{flushright}
\hfill
\vspace{18pt}
\begin{center}
{\Large \textbf{Warping the K\"ahler potential}}\\

\smallskip

{\Large \textbf{of  F-theory/IIB flux compactifications}}
\end{center}

\vspace{6pt}
\begin{center}
{\textsl{Luca Martucci}}

\vspace{1cm}
\textit{Dipartimento di Fisica ed Astronomia ``Galileo Galilei",  Universit\`a di Padova\\
\& I.N.F.N. Sezione di Padova,
Via Marzolo 8, 35131 Padova, Italy} \\  \vspace{6pt}
\end{center}

\vspace{12pt}

\begin{center}
\textbf{Abstract}

\end{center}

\vspace{4pt} {\small
\noindent 
We identify the low-energy K\"ahler potential of warped F-theory/IIB flux compactifications
whose light spectrum includes universal, K\"ahler, axionic and mobile D3-brane moduli.  
The derivation is based on four-dimensional local superconformal symmetry and holomorphy of brane instanton contributions and it 
reproduces and generalises previous partial results.
We compute the resulting kinetic terms, which show their explicit dependence on the warping. The K\"ahler potential satisfies the no-scale condition
 and produces,  at leading order in the large volume expansion, a specific correction to the unwarped K\"ahler potential.

\noindent }

\vspace{1cm}

%\noindent {\em Possible comment ..........................................................................................................................................}

\thispagestyle{empty}

%\vfill
%\vskip 5.mm
%\hrule width 5.cm
%\vskip 2.mm
%{\scriptsize
%e-mails: {\tt massimo.bianchi@roma2.infn.it, andres.collinucci@physik.uni-muenchen.de, luca.martucci@roma2.infn.it
%}}

\newpage

\setcounter{footnote}{0}

\tableofcontents
%
%\newpage

%%%%%%%%%%%%%%%%%%%%%%%%%%%%%%%%%%%%%%%%
\section{Introduction and summary}

Warped F-theory/IIB flux compactifications \cite{Becker:1996gj,Dasgupta:1999ss,Grana:2000jj,Gubser:2000vg,GKP,Grana:2001xn,Becker:2001pm} play a prominent  role in several phenomenologically inspired string constructions, for reviews see for instance \cite{Denef:2008wq,Maharana:2012tu}. In particular, the flux allows to stabilise  several moduli at tree-level and at the same time back-reacts on the geometry by generating a non-trivial warping, which can lead to interesting physical effects, as discussed e.g.\ in \cite{GKP,DeWolfe:2002nn,Kachru:2003aw,Kachru:2003sx,Giddings:2005ff,Frey:2006wv,Burgess:2006mn,Marchesano:2008rg,Marchesano:2010bs,Grimm:2012rg}. 

However, a systematic understanding of the low-energy effective  theory  of such flux compactifications is, so far, still missing.  The most common approach is to use the effective theory obtained from direct dimensional reduction by neglecting the back-reaction of  fluxes and other sources, approximating the warping to a constant \cite{Grana:2003ek,Grimm:2004uq,Jockers:2004yj,Jockers:2005zy,Grimm:2010ks}. On the other hand, attempts to repeat  a similar direct dimensional reduction by including a non-trivial warping, see for instance  \cite{DeWolfe:2002nn,Giddings:2005ff,Shiu:2008ry,Douglas:2008jx,Frey:2008xw,Chen:2009zi,Underwood:2010pm,Frey:2013bha}, are technically quite involved. These complications  obscure the $N=1$ supersymmetric structure, which should eventually show up in the effective theory. Indeed,  conclusive general answers have been  so far reached only in few  particular sub-cases.   

In this paper, following \cite{eff1,eff2}, we use an alternative strategy to identify the effective K\"ahler potential of warped F-theory/IIB compactifications. This strategy is based on the interplay between ten-dimensional geometry, four-dimensional local superconformal symmetry and holomorphy of brane instantons. In section \ref{sec:back}  we review the structure of the warped F-theory/IIB backgrounds and in section \ref{sec:KKK} we adapt the strategy of \cite{eff1,eff2}  to them.  
As it happens in the constant warping approximation, in general the K\"ahler potential can  only  be defined implicitly. We omit  from our discussion  the moduli associated with the axio-dilaton, the complex structure and the position of seven-branes, which typically get a mass from the flux-induced tree-level superpotential.  Hence, in section \ref{sec:universal}, we argue that the implicit K\"ahler potential has the very simple form
\be\label{intK}
K=-3\log \hat a+\text{const}
\ee
where $\hat a$ is a specific parametrisation  of  the universal modulus. The formula (\ref{intK}) is the warped counterpart of the analogous implicit  formula $K_{\rm unwarped}\simeq-2\log V$, where $V$ is the internal volume, obtained in the constant warping approximation \cite{Grana:2003ek,Grimm:2004uq,Jockers:2004yj,Jockers:2005zy,Grimm:2010ks}.  

In order to make (\ref{intK}) explicit, one should know how $\hat a$ depends on the chiral fields $\varphi^A$ which properly parametrise the scalar sector of the $N=1$ four-dimensional effective theory. This issue is addressed in section \ref{sec:kahler}, focusing on the case in which only the universal, K\"ahler and mobile D3-brane moduli are present. 
By using probe supersymmetric brane instantons, we introduce a natural set of chiral fields  $\varphi^A$ and  identify the relevant  dependence thereof on the background moduli. 
This implicitly defines $\hat a=\hat a(\varphi,\bar\varphi) $ and then, from  (\ref{intK}), $K=K(\varphi,\bar\varphi)$. 

As we discuss in section \ref{sec:kinetic}, even though the K\"ahler potential (\ref{intK}) is defined only implicitly, it is actually possible to explicitly compute the associated  low-energy kinetic terms. Furthermore,  we show that the  K\"ahler potential (\ref{intK}) is of no-scale type \cite{Cremmer:1983bf,Ellis:1983sf}.  
This result is actually required by the physical consistency of the theory and then  provides a strong consistency check.

Our results exhibit  the precise contribution of mobile D3-branes, background fluxes and the remaining D3-charge sources to  the effective K\"ahler potential.  
The contribution of D3-branes had been already considered before under certain restrictions or approximations \cite{DeWolfe:2002nn,Grana:2003ek,Giddings:2005ff,Baumann:2006th,Chen:2009zi}. On the other hand, to best of our knowledge,  the contribution of fluxes is completely new. As discussed in section \ref{sec:large},
by considering the large universal modulus limit, one can make more manifest the warping-induced correction to the  K\"ahler potential with respect to the unwarped case.
In particular, we explicitly  identify the leading correction as a function of the K\"ahler moduli, the D3-brane positions, the fluxes and the other D3-charge sources.

In section \ref{sec:axions}, going to a  weakly-coupled regime for technical simplicity, 
we show how one can adopt the same approach to incorporate the chiral fields associated with purely axionic moduli, neglected so far.
All the steps work similarly as in the previous sections. In particular, one can compute the explicit form of the kinetic terms and verify the no-scale condition.    

Finally, appendix A  contains some technical details and appendix B describes the application of the general results of the paper to a simple concrete class of $N=1$ 
flux compactifications on $T^6/\mathbb{Z}_2$ \cite{Kachru:2002he}, for which the K\"ahler potential can be made fully explicit.

In this paper we do not explicitly consider the incorporation of other sectors of the effective theory, as for instance the gauge sector and the chiral matter, and
the implications of higher-order corrections to the leading ten-dimensional supergravity theory. 
The investigation of these aspects  is left to the future.

 %%%%%%%%%%%%%%%%%%%%%%%%%%%%%%%%%%%%%%%%%

\section{Background structure}
\label{sec:back}

Let us review the structure of warped F-theory/IIB compactifications, by using  the type IIB description of these vacua  
\cite{Grana:2000jj,Gubser:2000vg,GKP,Grana:2001xn}. These vacua have an Einstein frame metric of the form
\be\label{metricansatz}
\d s^2_{10}=e^{2D}\d s^2_{\mathbb{R}^{1,3}}+e^{-2D}\d s^2_X
\ee
where $\d s^2_{\mathbb{R}^{1,3}}$ is the flat Minkowski four-dimensional metric,  the internal space $X$ is  K\"ahler   and the warping  $e^{2D}$ is generically non-constant along the internal directions. Furthermore for F-theory backgrounds, namely in presence of 7-branes,  the axio-dilaton $\tau\equiv C_0+\ii e^{-\phi}$ varies holomorphically along $X$ and is allowed to undergo non-trivial SL(2,$\mathbb{Z}$) monodromies.  The Ricci tensor $R^X$ associated with the internal K\"ahler metric $\d s^2_X$ is related to the dilaton $\phi$ by the formula:
\be\label{ricci}
R^X_{i\bar\jmath}=\del_i\del_{\bar\jmath}\phi
\ee

In addition, there can be a non-trivial three-form flux  $G_3=F_3-\ii e^{-\phi}H_3$, with $F_3=\d C_2-C_0H_3$, which must be imaginary-self-dual (ISD) on $X$, $*_XG_3=\ii G_3$. In other words $G_3$ can have non-vanishing primitive $(2,1)$ or $(0,3)$ components. Supersymmetry, which we will mostly assume, requires the $(0,3)$ component to be absent. 
In addition there can be supersymmetric world-volume fluxes supported on the world-volume of the seven branes. Furthermore, 
the R-R  $F_5$ is also non-trivial and its form is directly linked to the warping by
\be\label{F5}
F_5=*_X\d e^{-4D}+\d x^{0123}\wedge \d e^{4D}
\ee
where $*_X$ is computed by using the K\"ahler metric $\d s^2_X$. 

The $F_5$ Bianchi identity is
\be\label{F5BI}
\d F_5-H_3\wedge F_3=-Q^{\rm loc}_6
\ee
where $Q^{\rm loc}_6$ denotes the localised D3-charge current 
\be
Q^{\rm loc}_6=\ell^4_{\rm s}\sum_{I\in \text{D3's}}\delta^6_{I}-\frac14\ell^4_{\rm s}\sum_{O\in\text{O3's}}\delta^6_{O}+\ldots
\ee
We have introduced the string length $\ell_{\rm s}\equiv2\pi\sqrt{\alpha'}$ and  the six-form currents $\delta^6_{I}$ and $\delta^6_O$ localised at the position of the  $I$-th D3-brane and of the $O$-th O3-plane respectively.\footnote{\label{foot:current}In general, the $p$-form current $ \delta^p_{\Sigma}$ associated with a $6-p$ surface $\Sigma$ is such that, for any $(6-p)$-form $\alpha$ on $X$, $\int_{\Sigma}\alpha=\int_X\alpha\wedge  \delta^p_{\Sigma}$.}
In $Q^{\rm loc}_6$ the  ellipses  denote  other localised sources of D3-brane charge, 
induced by fluxes and curvature corrections on the 7-branes. 
By combining (\ref{F5}) and (\ref{F5BI}) one obtains the following equation for  the warp factor:
\be\label{weq}
\Delta\,e^{-4D}=\frac{1}{2\Im \tau}G_3\cdot \bar G_3+*_XQ^{\rm loc}_6
\ee
where we have introduced the Laplacian $\Delta\equiv -\nabla_m\nabla^m$.

The supersymmetric structure of these backgrounds is characterised by a Weyl Killing spinor $\epsilon$ of the form
\be\label{spinansatz}
\epsilon=e^{\frac{D}{2}}\zeta_{\rm R}\otimes \eta
\ee
where $\eta$ is an internal Weyl spinor satisfying the normalisation condition $\eta^\dagger\eta=1$, and   $\zeta_{\rm R}$ is an external right-handed spinor. 
The spinor $\eta$ can be used to construct the holomorphic  $(3,0)$-form $\Omega$ and the K\"ahler form $J$  as follows
\be\label{defJ}
J_{mn}=-\ii\eta^\dagger\gamma_{mn}\eta\quad~~~~~~\Omega_{mnp}=e^{\frac{\phi}{2}}\eta^T\gamma_{mnp}\eta
\ee
These satisfy the normalisation condition 
\be
\frac{1}{3!}J\wedge J\wedge J=\frac\ii 8\, e^{-\phi}\Omega\wedge \bar\Omega=\d{\rm vol}_X
\ee
where the volume form $\d {\rm vol}_X$ is computed by using the internal K\"ahler metric $\d s^2_X$.\footnote{In presence of 7-branes, $\Omega$ cannot be considered as a section of the canonical bundle $\calk_X$. Rather, it is a section of $\calk_X\otimes \call_Q$, where $\call_Q$ is a holomorphic line bundle associated with the non-trivial holomorphic  $\tau$. Supersymmetry requires that $\call_Q\simeq \calk_X^{-1}$ so that $\Omega$ is indeed globally defined and no-where vanishing.}

It is important to observe that this description of the background is intrinsically redundant. Indeed, one can always perform a constant rescaling of the four-dimensional metric 
\be\label{weylmetric}
\d s^2_{\mathbb{R}^{1,3}}\rightarrow e^{-2\omega}\d s^2_{\mathbb{R}^{1,3}}
\ee if accompanied by $e^{2D}\rightarrow e^{2D+2\omega}$ and $\d s^2_X\rightarrow e^{2\omega}\d s^2_X$.
Furthermore, in the spinorial ansatz (\ref{spinansatz}) one can perform a constant phase change $\zeta_{\rm R}\rightarrow e^{\frac{\ii}{2}\alpha}\zeta_{\rm R}$, if accompanied by $\eta\rightarrow e^{-\frac{\ii}{2}\alpha}\eta$. Under such reparametrizations $\Omega$ transforms as
\be\label{weylomega}
\Omega\rightarrow e^{3\omega-\ii\alpha}\Omega
\ee
 while  $J\rightarrow e^{2\omega} J$.

%%%%%%%%%%%%%%%%%%%%%%%%%%%%%%%%%%%%%%%%%%%%%%%%%%%%%%%%%%%%

\section{K\"ahler potential from warped volume}
\label{sec:KKK}

In this section we discuss how the K\"ahler potential  for warped F-theory/IIB compactifications is directly related to the warped volume of the internal space.
The K\"ahler potential that we obtain has the same  form of the K\"ahler potential proposed in \cite{DeWolfe:2002nn,Giddings:2005ff}.   
Our argument is based on local superconformal symmetry as in \cite{eff1,eff2}. It  is morally very close  to  arguments presented in other papers, as for instance in \cite{Haack:1999zv,Haack:2001jz,Grimm:2010ks,Grimm:2013bha,Junghans:2014zla}, in which  the K\"ahler potential is also directly related to  the volume-like pre-factor appearing in the dimensionally reduced Einsten-Hilbert term. On the other hand, the local superconformal symmetry provides a precise scheme in which  this conclusion is reached in a  neat and natural way.

\subsection{The conformal K\"ahler potential}

The backgrounds described in section \ref{sec:back} are characterised by several moduli, which in this section we collectively denote by $u^a$. 
In order to obtain the four-dimensional effective theory
one should allow for various components of the ten-dimensional ansatz to be slowly varying functions  of the external coordinates. In particular, the external flat metric $\d s^2_{\mathbb{R}^{1,3}}$ and the moduli must be promoted to  dynamical fields $g_{\mu\nu}(x)\d x^\mu\d x^\nu$ and  $u^a(x)$ respectively. Then, the vacuum metric ansatz  (\ref{metricansatz}) must be generalised to
\be\label{metrica}
\d s^2_{10}=e^{2D(y;u)}g_{\mu\nu}\d x^\mu\d x^\nu+e^{-2D(y;u)}\d s^2_{X}(y;u)+...
\ee
We have shown only the terms that correspond to the naive generalisation of  (\ref{metricansatz}). In fact, as discussed for instance in \cite{Gray:2003vw,Giddings:2005ff,Douglas:2008jx,Frey:2008xw}, 
in presence of warping these two terms do not constitute a consistent KK ansatz by themselves,  but must be supplemented by additional compensating terms (often referred to as `compensators'), which depend on the derivatives of the space-time dependent moduli $\partial_\mu u^a(x)$. These terms should then fill in the ellipses appearing on the right-hand side of (\ref{metrica}).

Now, in presence of warping, a standard dimensional reduction is generically quite cumbersome, in particular because of the compensating terms. 
However, since the vacua we start from are supersymmetric, one already knows that the four-dimensional  theory must be supersymmetric. Furthermore, crucially, in the dynamical ansatz (\ref{metrica}), the scaling symmetry (\ref{weylmetric}) is promoted to a gauge symmetry under arbitrary Weyl transformations $g_{\mu\nu}\rightarrow e^{-2\omega(x)}g_{\mu\nu}$.
An analogous conclusion holds for the arbitrary phase transformation of the spinors entering the dynamical generalisation of (\ref{spinansatz}).
Hence, in particular (\ref{weylomega}) is generalised to 
\be\label{weylgauge}
\Omega\rightarrow e^{3\omega(x)-\ii\alpha(x)}\Omega
\ee
We then see that the effective four-dimensional theory must be  gauge invariant under such complexified Weyl transformation. This tells us that the effective theory can be naturally regarded as  a superconformal supergravity. By definition, fields transforming by a phase $e^{w\omega(x)+\ii c\alpha(x)}$ under the above Weyl-chiral  symmetry are said to have Weyl-chiral weight $(w,c)$. Hence $\Omega$ has Weyl-chiral weight $(3,-1)$. We refer to \cite{Kallosh:2000ve} for an exhaustive discussion on $N=1$ superconformal supergravities. 

One of the nice features of superconformal supergravity is that the  Einstein-Hilbert term as well as the scalars' kinetic terms all derive from a D-term of the schematic form 
\be\label{Dterm}
-3\int\d^4\theta\, \caln(\Phi,\bar\Phi)
\ee
Here $\Phi^I$ denote the conformal chiral multiplets and we  refer to $\caln(\Phi,\bar\Phi)$ as the conformal K\"ahler potential. Consistency requires $\caln(\Phi,\bar\Phi)$ to have Weyl-chiral weight $(2,0)$.

 A discussion of the precise meaning of (\ref{Dterm}) can be found in \cite{Kallosh:2000ve}. Here we just need few bosonic terms which appear once (\ref{Dterm}) is written in components: 
\be\label{effsc}
\frac12\caln R_4 +3\,\caln_{I\bar J}\,\del_\mu\Phi^I\del^\mu\bar \Phi^{\bar J}+\ldots
\ee 
where $\caln_{I\bar J}(\Phi,\bar\Phi)\equiv \del_I\delbar_{\bar J}\caln(\Phi,\bar\Phi)$ and we use the same symbol for chiral multiplets as well as for their scalar components.

By appropriately performing the KK reduction one should be able to recover the terms appearing in (\ref{effsc}). Of course, in order to reconstruct the scalars' kinetic terms  one should properly take into account  the compensating terms, which contain the derivatives of the moduli $\del_\mu u^a$. 
However, there is a clear short-cut.  The scalars' kinetic terms are defined by the conformal K\"ahler potential $\caln(\Phi,\bar\Phi)$, which also appears as the prefactor of the Einstein term in (\ref{effsc}). Since such term  does not  contain space-time derivatives of the chiral fields, one can obtain $\caln$ by plugging the naive ansatz (\ref{metrica}) in the Einstein term of the ten-dimensional type IIB action 
\be
\frac{2\pi}{\ell^8_{\rm s}}\int \sqrt{-g_{10}}\, R_{\rm 10}\, ,
\ee
simply ignoring the problematic compensating terms. In this way, one easily gets 
\be\label{implN}
\caln=\frac{4\pi}{\ell^8_{\rm s}}\int_X e^{-4D}\d {\rm vol}_X
\ee
Of course, this is not the end of the story, since (\ref{implN}) provides only a rather implicit formula for $\caln(\Phi,\bar\Phi)$.\footnote{In \cite{eff1} a holomorphic parametrisation of the generic (massless and massive) fluctuations was introduced, which used the formalism of generalised complex geometry. By using this parametrisation, it was shown how the conformal K\"ahler potential (\ref{implN}) and an associated GVW-like  superpotential \cite{Gukov:1999ya} reproduce the {\em complete} set  of ten-dimensional equations of general $N=1$ compactifications to ${\rm Mink}_4$ and AdS$_4$ found in \cite{gmpt}.}

\subsection{The conformal compensator and the K\"ahler potential}

Part of the possible background moduli correspond to the deformations of the axio-dilaton $\tau$, which encodes  the value of the string coupling as well as the seven-brane positions, and of the holomorphic (3,0) form $\Omega$, which specifies  the complex structure of the K\"ahler space $X$. The dual M-theory perspective elegantly unify these data into the holomorphic $(4,0)$ form associated with the dual elliptically fibered Calabi-Yau four-fold. Although these moduli could in principle be included in the discussion, we assume them to be stuck at a given value. This condition can be indeed dynamically  enforced by the presence of fluxes. This means that $\tau$ and $\Omega$ are fixed, up to the latter's overall normalisation, which cannot be fixed basically because of the gauge invariance    (\ref{weylgauge}). It is then convenient to isolate such degree of freedom by writing
\be\label{rescomega}
\Omega\equiv \ell^6_{\rm s}\,Y^3 \Omega_0
\ee
where $Y$ is a completely arbitrary complex parameter which transforms with Weyl-chiral weight $(1,-\frac13)$
 \be
 Y\ \rightarrow\  e^{\omega(x)-\frac{\ii}3\alpha(x)} Y
 \ee 
while $\Omega_0$ is some fiducial dimensionless holomorphic $(3,0)$ form, satisfying a normalisation condition
\be\label{norm}
\frac{\ii}{8}\int_Xe^{-\phi}\Omega_0\wedge\bar\Omega_0={\rm v}_0
\ee 
with ${\rm v}_0$ being a dimensionless constant.\footnote{The factor $e^{-\phi}$ is necessary to make the integrand  a proper volume-form even in presence of SL(2,$\mathbb{Z}$) monodromies.} For instance, one may set ${\rm v}_0=1$.

The complex parameter $Y$ appears as a chiral field in the low-energy superconformal effective theory and can be used as conformal compensator to gauge fix the theory to a standard 
Poincar\'e supergravity. A detailed description of the procedure can be found  in \cite{Kallosh:2000ve} and we will be sketchy.  

First, one can isolate the dependence of  $\caln$ on the conformal compensator by writing it as
\be\label{Nsplit}
\caln=|Y|^2e^{-\frac{1}{3}K(\varphi,\bar \varphi)}
\ee
where  $\varphi^A$  denote chiral fields of Weyl-chiral weight $(0,0)$ parametrising the background moduli. Then, one can gauge-fix the superconformal symmetry by imposing 
\be\label{scgauge}
Y= M_{\rm P}\,e^{\frac{1}{6}K(\varphi,\bar\varphi)}
\ee 
This implies that $\caln\equiv M^2_{\rm P}$ and the first term in (\ref{effsc}) reduces to the standard Einstein term $\frac12 M^2_{\rm P} R_4$. 
On the other hand, after the gauge fixing (\ref{scgauge}) the  kinetic terms appearing in (\ref{effsc}) produce the more familiar kinetic terms
\be\label{Ekin}
-M^2_{\rm P}\,K_{A\bar B}(\varphi,\bar \varphi)\,\del_\mu \varphi^A\del^\mu\bar\varphi^{\bar B}
\ee 
where $K_{A\bar B}\equiv \del_A\delbar_{\bar B}K$. Hence the function $K(\varphi,\bar \varphi)$ introduced in (\ref{Nsplit}) can  be identified with the K\"ahler potential of standard $N=1$ supergravity. 

In order to identify the ten-dimensional expression for $K$ we observe that the redefinition  (\ref{rescomega}) is naturally associated  with the redefinitions $e^{2D}\equiv \ell^2_{\rm s}|Y|^2\,e^{2A}$, $\d s^2_{X}\equiv\ell^4_{\rm s}|Y|^2\d s^2_{X,0}$ and  $J\equiv\ell^4_{\rm s}|Y|^2J_0$, so that the ten-dimensional metric now reads
\be\label{rescmetric}
\d s^2_{10}=\ell^2_{\rm s}e^{2A}|Y|^2\d s^2_4+\ell^2_{\rm s}e^{-2A}\d s^2_{X,0}
\ee
where the metric $\d s^2_{X,0}$ still satisfies (\ref{ricci}).
Notice that $\d{\rm vol}_{X,0}\equiv \frac{1}{3!}J_0\wedge J_0\wedge J_0$
can be considered as  a fiducial  volume-form with normalisation fixed by (\ref{norm}):  
\be\label{norm2}
\int_X\d{\rm vol}_{X,0}=\frac{1}{3!}\int_X J_0\wedge J_0\wedge J_0={\rm v}_0
\ee
In practice, this normalisation condition removes the overall rescaling from the set of possible deformations of the K\"ahler form $J_0$. Such rescaling parametrises  the universal modulus in standard unwarped compactifications. On the other hand, as we will review in the next section, in warped backgrounds the universal modulus is encoded in the warp factor and then no physical degree of freedom is lost  in the normalisation condition  (\ref{norm2}).

We can now write the (implicit) form of the K\"ahler potential $K(\varphi,\bar\varphi)$. From (\ref{Nsplit}) and the above field redefinitions one obtains
 \be\label{wkahler}
 \begin{aligned}
K(\varphi,\bar\varphi)&=-3\log\Big(4\pi\int_Xe^{-4A}\d{\rm vol}_{X,0}\Big)\\
&\equiv -3\log\Big(\frac{1}{3!}\int_Xe^{-4A}J_0\wedge J_0\wedge J_0\Big)-3\log{4\pi}
\end{aligned}
\ee
This K\"ahler potential has the same form as the K\"ahler potential proposed in \cite{DeWolfe:2002nn,Giddings:2005ff}.

\section{Universal modulus and K\"ahler potential}
\label{sec:universal}

In this section we  show that the K\"ahler potential (\ref{wkahler}) admits a simple expression in terms  of the universal modulus.  
So, we start by reviewing the origin  of the universal modulus in warped F-theory/IIB compactifications. 
First notice that the rescaled warp-factor $e^{2A}$ introduced in the previous section satisfies  an equation formally identical to (\ref{weq}),
\be\label{weq0}
\Delta_0\, e^{-4A}=\frac{1}{\ell^4_{\rm s}}*_0Q_6\equiv \frac{1}{\ell^4_{\rm s}}\Big(\frac{1}{2\Im \tau}G_3\cdot \bar G_3+*_0Q^{\rm loc}_6\Big)
\ee
where all quantities are now computed by using the rescaled metric $\d s^2_{X,0}$. 

The universal modulus  is readily identified by looking at (\ref{weq0}) and noticing that it completely determines $e^{-4A}$ up to an additive constant.  
One can identify  the universal modulus $a$ with such a constant, by writing 
\be\label{splitwarp}
e^{-4A}=a+e^{-4A_0}
\ee
where $e^{-4A_0}$ is a particular solution of (\ref{weq0}).  
Notice that the split (\ref{splitwarp}) is not unique as one could shift $a\rightarrow a+c$ and $e^{-4A_0}\rightarrow e^{-4A_0}-c$. One can fix 
such redundancy by imposing
\be\label{wgauge}
\int_Xe^{-4A_0}\d{\rm vol}_{X,0}={\rm v}_0^{\rm w}
\ee
for an arbitrary constant  ${\rm v}_0^{\rm w}$.
Hence, recalling (\ref{norm2}), we can write the K\"ahler potential (\ref{wkahler})  as follows
\be\label{simpk}
K=-3\log ({\rm v}_0\, a+ {\rm v}_0^{\rm w})-3\log 4\pi
\ee

Consider now the particular solution  $e^{-4\hat A_0}$  of (\ref{weq0}) provided  by the Green's operator of the Laplacian $\Delta_0$. The associated Green's function
$G(y;y')$ satisfies the equation
\be\label{green}
\Delta_{0,y}G(y;y')=*_0\delta^6_{y}(y')-\frac{1}{{\rm v}_0}
\ee 
The particular solution $e^{-4\hat A_0}$ is then given by
\be\label{warpgreen}
e^{-4\hat A_0(y)}=\frac{1}{\ell^4_{\rm s}}\int_XG(y;y')Q_6(y')
\ee
By definition,  the Green's operator vanishes on constant functions and maps coexact functions to coexact functions, see for instance \cite{Warner}. 
This implies that  $e^{-4\hat A_0(y)}$ defined by (\ref{warpgreen}) is a coexact function on $X$, so that
\be\label{norm0}
\hat{\rm v}_0^{\rm w}\equiv\int_X e^{-4\hat A_0} \d{\rm vol}_{X,0}=\frac1{3!}\int_X e^{-4\hat A_0} J_0\wedge J_0\wedge J_0\, \equiv 0
\ee 

The particular solution (\ref{warpgreen}) is associated with  a particular parametrisation $\hat a$ of the universal modulus, defined by
  \be\label{partsplit}
 e^{-4A}=\hat a+e^{-4\hat A_0}
 \ee
By using this specific split of $e^{-4A}$ and (\ref{norm0}), the K\"ahler potential (\ref{wkahler}) takes the form
\be\label{partk}
K=-3\log ({\rm v}_0\, \hat a)-3\log 4\pi
\ee

We then see that the K\"ahler potential (\ref{wkahler}) reduces to a simple function of just the universal modulus $\hat a$ (or $a$). 
On the other hand, the formula (\ref{partk}) (or  (\ref{simpk})) is still implicit. In order to make it explicit we  need to understand how the universal modulus  $\hat a$  
(or $ a$) is parametrised in terms of appropriate holomorphic coordinates $\varphi^A$. This issue will be addressed in the following sections.

Notice that we have been careful in not dropping out additional constants  in $K$, which are  irrelevant in the computation of the K\"ahler metric. Nevertheless,  such terms could be relevant for the computation of other quantities, as the potential 
\be\label{pot4D}
V_{\rm 4D}=\frac{1}{M^2_{\rm P}}e^K\left(K^{A\bar B}D_AWD_{\bar B}\bar W-3|W|^2\right)
\ee
which depends on the normalisation of the superpotential $W$ as well. For instance,  with our choice of the K\"ahler potential, the classical GVW superpotential  \cite{Gukov:1999ya,GKP}  has the following specific normalisation \cite{eff1} 
\be\label{GVW}
W_{\rm tree}=\frac{\pi M^3_{\rm P}}{\ell^2_{\rm s}}\int_X\Omega_0\wedge G_3
\ee
For later convenience, let us recall that (\ref{GVW}) depends just on the axio-dilaton, complex-structure moduli and seven-brane moduli, and is at the origin of their  stabilisation, which we assume. Hence, in the present paper $W_{\rm tree}$  can be considered as constant, $W_{\rm tree}\equiv W_0$.   For supersymmetric vacua, i.e.\ with $(2,1)$ and primitive $G_3$, $W_0=0$. On the other hand, as reviewed in section \ref{sec:back},   supersymmetry is broken if the ISD $G_3$ has a non-vanishing $(0,3)$ component, still preserving the ten-dimensional equations of motion. In this case $W_0\neq 0$.

%%%%%%%%%%%%%%%%%%%%%%%%%%%%%%%%%%%%%%%%%%%%%%%%%%%%%%
%%%%%%%%%%%%%%%%%%%%%%%%%%%%%%%%%%%%%%%%%%%%%%%%%%%%%%%

\section{Decoding the K\"ahler potential}
\label{sec:kahler}

 In order to interpret (\ref{partk}), or (\ref{simpk}), as a   K\"ahler potential, it is necessary to identify an appropriate  holomorphic parametrisation $\varphi^A$ of the background moduli. A  key  observation is that the background can be `probed' by supersymmetric Euclidean D3-branes (or vertical M5-branes in the dual M-theory description). It is well known that these branes  enter the path-integral as instantonic corrections and can generate new F-terms in the effective action. 
These terms contain a universal factor $e^{- S_{\rm D3}}$, where $S_{\rm D3}$ is the Euclidean on-shell D3-brane action. Crucially, by four-dimensional supersymmetry, $S_{\rm D3}$ must depend holomorphically on the chiral fields $\varphi^A$.\footnote{Notice that the following discussion can be straightforwardly applied  to seven-brane four-dimensional   gauge-couplings as well. Indeed,  by dimensionally reducing  the effective action of a  D7-branes (or, by SL(2,$\mathbb{Z}$) duality, of a more general seven-brane) wrapped on a divisor $D$, one readily gets the complexified gauge coupling $\tau_{\rm YM}=\frac{1}{2\pi}S_{\rm D3}$.}

Now, the bosonic Euclidean D3-brane action can be written as 
\be\label{E3action}
S_{\rm D3}=S_{\rm DBI}-\ii S_{\rm CS}
\ee 
where $S_{\rm DBI}$ and $S_{\rm CS}$ are the standard Dirac-Born-Infeld (DBI) and Chern-Simons (CS) terms. In fact, in presence of warping and fluxes, the definition of $S_{\rm CS}$
is problematic, since the R-R potentials are not globally defined, see for instance \cite{Witten:1999eg} for a discussion in the analogous case of heterotic compactifications. 

On the other hand, $S_{\rm DBI}$ has no such ambiguities and one can in principle compute its dependence on the background moduli.  
At the same time $S_{\rm DBI}$ must be the real part of a chiral field. Hence, it can be used to identify the dependence  of the real part  of a set of chiral fields on the background moduli.  As we will see, this is sufficient to implicitly define $\hat a$, and then the K\"ahler potential  (\ref{partk}), as functions of chiral fields.

 In this section we apply this strategy by assuming that no two-form axionic moduli and seven-brane Wilson lines are present. This condition is better expressed in the dual elliptically fibered Calabi-Yau four-fold $\hat X$, by requiring that   $b^3(\hat X)-b^3(X)=0$, see e.g.\ \cite{Grimm:2010ks}.   The inclusion of axions will be discussed in section \ref{sec:axions}.
 
 Hence, in addition to the universal modulus $\hat a$ discussed in section \ref{sec:universal}, we allow for three other   kinds of moduli. The $h^{1,1}(X)-1$ K\"ahler moduli  describing the deformations of $J_0$ that preserve the normalisation condition (\ref{norm2}),\footnote{We are implicitly assuming that the proper orientifold  projections are satisfied and we are ignoring for simplicity the possible lift of K\"ahler moduli induced by flux primitivity conditions. These aspects can be easily  taken into account, as in the concrete example on $T^6/\mathbb{Z}_2$ discussed in appendix \ref{app:tori}.}  the $h^{2,2}(X)=h^{1,1}(X)$ moduli describing the $C_4$ axions, and the $6\times N_{\rm D3}$ (real) moduli  
 describing the positions of the $N_{\rm D3}$ D3-branes in the internal space.
  The latter have a natural holomorphic parametrisation provided by the bulk complex coordinates $z^i$, $i=1,2,3$. Hence, we denote the chiral fields describing the D3-brane positions by $Z^i_I$, with $I=1,\ldots,N_{\rm D3}$. We then need to understand how the remaining moduli organise into chiral fields.

Let us introduce a set of integral closed two-forms $\omega^a$, whose associated cohomology classes $[\omega_a]$ provide a basis of $H^2(X;\mathbb{Z})$, and expand  the K\"ahler form $J_0$ in cohomology as follows
\be\label{cohoJ}
[J_0]=v_a\, [\omega^a]
\ee
Then the constraint (\ref{norm2}) can be written as
\be\label{constr}
\frac{1}{3!}\,v_av_b v_c\,\cali^{abc}={\rm v}_0
\ee
where we have introduced the intersection matrix
\be\label{intn}
\cali^{abc}\equiv\int_X\omega^a\wedge \omega^b\wedge \omega^c
\ee
The constraint (\ref{constr}) identifies $h^{1,1}-1$ K\"ahler moduli  out of the $h^{1,1}$ parameters $v_a$. Hence we will refer to $v_a$ as {\em constrained} K\"ahler moduli. 

Now,  a supersymmetric D3-brane must wrap a homographic four-cycle (an effective divisor) $D$, with anti-self-dual world-volume flux $\calf=\frac{1}{2\pi}\ell^2_{\rm s}F_{\rm D3}-B_2|_{D}$.   These conditions are equivalent to requiring that the D3-brane is calibrated in the generalised sense of \cite{luca1}, so  that the associated on-shell DBI action reduces to\footnote{See \cite{Bianchi:2011qh,Bianchi:2012kt,Martucci:2014ema} for more details  on the supersymmetric structure of D3-brane instantons in F-theory compactifications, in the IIB framework adopted in the present paper.}
\be\label{susyDBI}
\begin{aligned}
S_{\rm DBI}=\pi\int_{D}e^{-4A}J_0\wedge J_0-\frac{\pi}{\ell^4_{\rm s}}\int_{D}e^{-\phi}\calf\wedge \calf
\end{aligned}
\ee
Notice that the second term does not depend on the background moduli we are considering in this section. So, for the moment, it is a constant and  we can ignore it.

Now, there are $h^{1,1}(X)$ independent divisors $D^a$ and we choose the $[\omega^a]\in H^2(X;\mathbb{Z})$  to be Poincar\'e dual to such divisors:
\be
[\omega^a]={\rm PD}_X(D^a)
\ee
By wrapping   D3-branes along linear combinations of the divisors $D^a$, the associated $S_{\rm DBI}$'s can detect all geometric moduli, i.e.\ universal modulus as well as all  $h^{1,1}(X)-1$ constrained K\"ahler moduli $v_a$. On the other hand  $S_{\rm CS}$ would detect the $h^{1,1}(X)$ R-R axions, which should complexify the geometric moduli. However, since the K\"ahler potential (\ref{partk}) (or (\ref{simpk})) depends just on geometric moduli,  we do not actually need the information captured by $S_{\rm CS}$. 

We can then focus on the integrals
\be\label{Imoduli}
I^a\equiv\frac12\int_{D^a}e^{-4A}J_0\wedge J_0
\ee
According to the arguments presented above, these must correspond to $h^{1,1}(X)$ chiral fields $\rho^a$ through a relation of the form
\be\label{Imoduli2}
I^a=\Re\rho^a+(\text{hol}+\overline{\text{hol}})
\ee
where the $(\text{hol}+\overline{\text{hol}})$  contribution is the real part of some  holomorphic quantity depending only on the remaining chiral fields. 

Of course, deciding what to include in $(\text{hol}+\overline{\text{hol}})$ and what to absorb in the definition of $\Re\rho^a$ is a matter of choice. 
Indeed, possible different choices are related by a redefinition of the form $\rho^a\rightarrow \rho^a+\text{hol}$, which preserves the holomorphic parametrisation  of the chiral fields. 
In the following, our choice will be the minimal one. Namely, whenever we will be able to isolate a contribution to $I^a$ of the form $\text{hol}+\overline{\text{hol}}$  (not depending on $\hat a$ and $v_a$) we will omit it from the definition of $\Re\rho^a$.

\subsection{Decoding the moduli dependence}
\label{sec:decod}

In order to proceed, we need to decode the dependence of the integrals (\ref{Imoduli}) on the background moduli.
Let us  first use the decomposition (\ref{partsplit}) and write $I^a$ as
\be\label{Ia1}
I^a\equiv \hat a  \calv^a (v) +\frac12\int_{D^a}e^{-4\hat A_0}J_0\wedge J_0
\ee
where
\be
\calv^a(v)\equiv \frac{1}{2}\int_{D^a}J_0\wedge J_0=\frac{1}{2}\cali^{abc}v_bv_c
\ee
The effect of warping is encoded in the second term on the r.h.s.\ of (\ref{Ia1}), which can depend on the  constrained K\"ahler moduli $v_a$ and on the D3-brane 
 moduli $Z^i_I$. In order to learn more about such dependence  let us write the second term on the r.h.s.\ of (\ref{Ia1}) as
\be\label{intr}
\frac12\int_Xe^{-4\hat A_0}J_0\wedge J_0\wedge \delta^2_{D^a}=\int_Xe^{-4\hat A_0}(J_0\lrcorner \delta^2_{D^a})\d{\rm vol}_{X,0}
\ee
where $ \delta^2_{D^a}$ is the two-form current (cf.\ footnote \ref{foot:current}) associated with the divisor $D^a$  and 
\be
J_0\lrcorner \delta^2_{D^a}\equiv \frac12 J_0^{mn}(\delta^2_{D^a})_{mn}
\ee
Now, the localised term $J_0\lrcorner \delta^2_{D^a}$ admits a nice explicit realisation  in terms  of the section $\zeta^a(z)$ of the line bundle $\calo_X(D^a)$
whose vanishing locus identifies $D^a$: $D^a=\{\zeta^a=0\}$. Indeed,  the Poincar\'e-Lelong equation 
\be
\ii\del\delbar\Re(\log\zeta^a)=\pi\,\delta^2_{D^a}
\ee
implies that
\be\label{deltazeta}
J_0\lrcorner \delta^2_{D^a}=-\frac{1}{2\pi}\Delta_0\,  \Re\log\zeta^a(z)
\ee
Here we have used the fact that, since the internal space is K\"ahler, the Laplacian can be written as
\be
\Delta_0=-2g^{i\bar\jmath}\del_i\delbar_{\bar\jmath}=-2J_0\lrcorner(\ii\del\delbar)
\ee
By plugging (\ref{deltazeta}) into (\ref{intr}), one could be tempted to integrate by parts $\Delta_0$ and then use (\ref{weq0}). 
However, such procedure would not  be admissible, since $\zeta^a(z)$ is a 
 section of a non-trivial line bundle and then one would end with a  meaningless expression. 
 
 In order to find a way  out of this problem,
let us choose the integral two-forms  $\omega^a$ to be  $(1,1)$ and {\em harmonic}, 
which is always possible within their cohomology class. 
Since $J_0$ is harmonic, we can promote the cohomological identity (\ref{cohoJ}) to the pointwise identity
\be\label{Jexp}
J_0=v_a\omega^a
\ee
It is important to realise  that, in general, the $\omega^a$'s  actually depend on the constrained K\"ahler moduli $v_a$. 

Now, by the $\del\delbar$-lemma we can  introduce 
a set of $h^{1,1}(X)$ local  `potentials'  $\kappa^a(z,\bar z;v)$, which  generically depend also on the $v_a$'s, such that 
\be\label{defkappa}
\omega^a=\ii\del\delbar\kappa^a
\ee
Clearly the $\kappa^a$'s cannot be extended to global functions, since the $\omega^a$'s are non-trivial in cohomology.
Indeed,  $e^{-2\pi\kappa^a}$ can be identified with  a metric on the line bundle $\calo_X(D^a)$ associated with the divisor $D^a$, whose curvature is given by $2\pi\omega^a$, see e.g.\  \cite{GH}. 
More explicitly, let us consider a holomorphic transformation
\be
\zeta^a(z)\rightarrow e^{2\pi\chi^a(z)}\zeta^a(z)
\ee
 that  relates the expressions of $\zeta^a(z)$ on different local patches.
 Then,  $\kappa^a(z,\bar z;v)$ must correspondingly transform as
\be\label{genkt}
\kappa^a(z,\bar z;v)\rightarrow \kappa^a(z,\bar z;v)+ \chi^a(z)+\bar\chi^a(\bar z)
\ee
Hence, the combination 
\be\label{glofunct}
\pi\kappa^a(z,\bar z;v)-\Re\log\zeta^a(z)\equiv -\frac12\log e^{-2\pi\kappa^a}|\zeta^a|^2
\ee
is a globally well defined function on $X$.

Notice now that, since we have chosen the closed integral two-forms $\omega^a$ to be harmonic,  the functions $J_0\lrcorner \omega^a$ are harmonic too, i.e.\  they are  constant along $X$ (although  in general they depend on the constrained K\"ahler moduli $v_a$). By setting $d^a(v)\equiv J_0\lrcorner \omega^a$, we can rewrite (\ref{deltazeta}) as
\be\label{jdelta}
J_0\lrcorner \delta^2_{D^a}=\frac{1}{2\pi}\Delta_0  \Big[\pi\kappa^a(z,\bar z;v)-\Re\log\zeta^a(z) \Big]+ d^a(v)
\ee
We see that we have obtained a formula for $J_0\lrcorner \delta^2_{D^a}$ in terms of the globally well defined function (\ref{glofunct}).
By plugging (\ref{jdelta}) into (\ref{intr}), the contribution containing the constant $d^a(v)$ vanishes because of (\ref{norm0}), and  we are left with
\be
\frac1{2\pi}\int_Xe^{-4\hat A_0}\Delta_0  \big( \pi\kappa^a-\Re\log\zeta^a\big)\d{\rm vol}_{X,0}
\ee
Now we are legitimised to integrate  by parts the Laplacian and, by using (\ref{weq0}), we arrive at the following  remarkable identity 
\be\label{fincont}
\frac12\int_{D^a}e^{-4\hat A_0}J_0\wedge J_0=\frac1{2\pi\ell^4_{\rm s}}\int_X\big(\pi\kappa^a-\Re\log\zeta^a\big)Q_6
\ee
We recall that the six-form $Q_6$ denotes the D3-brane charge density and must satisfy the tadpole condition
\be\label{tadpolecond}
\int_X Q_6=0
\ee
Hence, even though both $\kappa^a$ and $\log\zeta^a$ are defined up to an additive constant, the identity (\ref{fincont}) as well as all results that will follow from it
are not affected by this ambiguity.    

One can isolate the contribution of the mobile D3-branes  to $Q_6$  by writing
\be\label{Qsplit}
Q_6\equiv \ell^4_{\rm s}\sum_{I}\delta^6_{I}+Q^{\rm bg}_6
\ee
where $Q^{\rm bg}_6$ is the D3-charge 6-form induced by bulk fluxes, O3-planes as well as other  localised D3-charge sources:
\be\label{Qbg}
Q^{\rm bg}_6=F_3\wedge H_3-\frac{1}{4}\ell^4_{\rm s}\sum_{O\in{\text{O3's}}}\delta^6_O+\ldots
\ee
By plugging (\ref{Qsplit}) into (\ref{fincont}) and inserting the result back in (\ref{Ia1}), we finally arrive at the following  more explicit form for $I^a$:
\be\label{Idec2}
I^a(v,Z,\bar Z)=\hat a\,\calv^a(v)+h^a(v)+\frac12\sum_I \kappa^a(Z_I,\bar Z_I;v)-\frac{1}{2\pi}\Re\log\Big[\prod_I\zeta^a(Z_I)\Big]
\ee
where\footnote{\label{foot:O3}Notice that $h^a(v)$ diverges to $-\infty$ if the divisor $D^a$ touches some O3-planes. This is a consequence 
of the localized negative tension associated with the O3-planes. This is expected to be an artefact of the effective supergravity description, which should be cured 
by   some higher-order or non-perturbative physical effect, as it happens for O7-branes in F-theory. In any case, one may judiciously choose the divisors $D^a$, in order to avoid this issue.} 
\be\label{defh}
h^a(v)\equiv \frac1{2\pi\ell^4_{\rm s}}\int_X\big(\pi\kappa^a-\Re\log\zeta^a \big)Q^{\rm bg}_6
\ee

We stress that $I^a$ depends  on the specific choice of the divisor $D^a$, since $h^a(v)$ depends on the associated holomorphic section $\zeta^a(z)$. The choice of a different divisor $\tilde D^a$,
homologous to $D^a$, would correspond to a different section $\tilde \zeta^a(z)$, and then to a different warping-induced contribution:
\be
\tilde I^a(v,Z,\bar Z)= I^a(v,Z,\bar Z)+\frac{1}{2\pi\ell^4_{\rm s}}\int_X \Re(\log\zeta^a-\log\tilde\zeta^a)Q_6
\ee
The second term on the r.h.s.\ contributes to the $(\text{hol}+\overline{\text{hol}})$ part in the prescription  (\ref{Imoduli2}) and can then be reabsorbed in a holomorphic redefinition of the chiral fields $\rho^a$.

\subsection{Chiral fields and K\"ahler potential}
\label{sec:chiral}

We can now come back to  the prescription  (\ref{Imoduli2}) to express the set of chiral fields $\rho^a$, or more precisely  their real part, in terms of the background moduli.
Indeed, by (\ref{Idec2})  we are naturally led to the following identification
\be\label{rhoinv}
\Re\rho^a=\hat a\,\calv^a(v)+h^a(v)+\frac1{2}\sum_I \kappa^a(Z_I,\bar Z_I;v)
\ee
Notice that, by consistency, under the transformations (\ref{genkt}) $\rho^a$ must transform as 
\be\label{rhotran}
\rho^a\rightarrow \rho^a +\sum_I \chi^a(Z_I)
\ee
Hence the chiral fields $\rho^a$ are non-trivially fibered over the moduli space $\calm_{\rm D3}$ of the D3-branes\footnote{ See for instance  \cite{Kachru:2003sx})} which can be identified with $N_{\rm D3}$ copies of the internal space $X$, modded out by the permutation group $S_{N_{\rm D3}}$. In other words, the total moduli space $\calm_{\rm tot}$ is given by a non-trivial fibration over $\calm_{\rm D3}$.

Now,  the equations (\ref{rhoinv}) implicitly define $\hat a$ and $v_a$ as functions of $\Re\rho^a$, $Z^i_I$ and $\bar Z^{\bar\imath}_I$. In particular, from  (\ref{partk}), one can in principle write
\be\label{complimpl}
K(\rho,\bar\rho,Z,\bar Z)=-3\log\big[ {\rm v}_0\,\hat a(\Re\rho,Z,\bar Z)\big]+ c_0
\ee
where 
\be
c_0=-3\log 4\pi
\ee
Of course, the main difficulty is the identification of the explicit form of the function $\hat a(\Re\rho,Z,\bar Z)$, for which unfortunately there is no general solution.
Simple cases in which one can be more explicit are discussed in the following. Nevertheless, as we will see in section \ref{sec:kinetic},
one can compute the effective kinetic terms  by using just (\ref{rhoinv}) and (\ref{complimpl}).

In order to make the relation with the literature on unwarped compactifications more immediate, one can combine $\hat a$ and the constrained K\"ahler moduli $ v_a$
into   the following $h^{1,1}(X)$ unconstrained moduli 
\be\label{vunconst}
\hat v_a=v_a\sqrt{a}\
\ee
These can be considered as moduli of an {\em auxiliary}   K\"ahler form $\hat J=\hat v_a \omega^a=\sqrt{a}\, J_0$, which would have a real geometrical interpretation only if one could consistently neglect $e^{-4\hat A_0}$ with respect to $\hat a$ in (\ref{partsplit}), i.e.\ only if the warping could be considered constant.

By using the moduli $\hat v_a$ the K\"ahler potential (\ref{complimpl}) can be written as
\be\label{unk}
K(\rho,\bar\rho,Z,\bar Z)=-2\log\Big(\frac{1}{3!}\cali^{abc}\hat v_a\hat v_b\hat v_c\Big)-\log {\rm v}_0+c_0
\ee
while the relations (\ref{rhoinv}) become
\be\label{rhoinvun}
\Re\rho^a=\frac12\cali^{abc}\hat v_b\hat v_c+h^a(\hat v)+\frac1{2}\sum_I \kappa^a(Z_I,\bar Z_I;\hat v)
\ee
Here we have used the fact that we can write $\kappa^a(z,\bar z;v)=\kappa^a(z,\bar z;\hat v)$ and $h^a(v)=h^a(\hat v)$, since the harmonicity condition on $\omega^a$ does not change if it is defined by $\hat J$ instead  of $J_0$. 

The description of the K\"ahler potential provided by (\ref{unk}) and (\ref{rhoinvun}) is formally closer to the one obtained in \cite{Grana:2003ek,Grimm:2004uq,Grimm:2010ks} in the constant warping approximation. 
Basically, the K\"ahler potential can be casted in the universal form (\ref{unk}), while the new ingredients like D3-branes and fluxes are 
completely encoded in the functional dependence $\hat v_a=\hat v_a(\Re\rho,Z,\bar Z)$ which should be obtained by inverting (\ref{rhoinvun}).
In particular, in \cite{Grana:2003ek} an analogous contribution of the D3-branes was obtained by using the probe D3-brane effective action expanded around a given point.
In our language, such contribution can be reproduced by expanding $Z^i_I\simeq Z^i_{\uu{0}I}+
\Phi^i_I$ and approximating $\kappa^a(Z_I,\bar Z_I;\hat v)\simeq -\ii \omega_{i\bar\jmath}(Z_{\uu{0}I},\bar Z_{\uu{0}I})\Phi_I^i\bar\Phi_I^{\bar\jmath}$.

On the other hand, to best of our knowledge, the contribution of the background fluxes and of the other localized D3-charge sources 
to the definition of the chiral coordinates had never been explicitly considered so far. 
This is completely encoded in the functions $h^a(v)\equiv h^a(\hat v)$. As we will see in section \ref{sec:kinetic}, the presence of such terms is  crucial in the derivation 
of the (warped) kinetic terms of the effective action. 

As a final comment, notice that the relations (\ref{rhoinv}), or (\ref{rhoinvun}), involve the background complex structure and, more implicitly,  the seven-brane positions,  whose  induced D3-charge  would contribute to $Q^{\rm bg}_6$. In this paper we are assuming the corresponding moduli to be frozen.
However, in any more complete treatment that includes (part of) these moduli as dynamical,  they would mix  with the remaining moduli in a non-trivial way, 
producing a non-diagonal K\"ahler potential, see also \cite{Grana:2003ek,eff1,eff2}.

%%%%%%%%%%%%%%%%%%%%%%%%%%%%%%%%%%%%%%%%%%%%%%%%%%%%%%%%
%%%%%%%%%%%%%%%%%%%%%%%%%%%%%%%%%%%%%%%%%%%%%%%%%%%%%%%%

\subsection{A comment on D3-brane instantons}
\label{sec:D3}

Consider a supersymmetric Euclidean D3-brane wrapping a divisor $D$, associated with a holomorphic section $\zeta_D(z)$. In homology, we can expand $D\simeq n_aD^a$.
Proceeding as in subsection \ref{sec:decod} and using (\ref{rhoinv}), one arrives at the identity
\be
\frac12\int_De^{-4A}J_0\wedge J_0= n_a \Re\rho^a-\frac{1}{2\pi}\Re\sum_I\log\zeta_D(Z_I)+\frac{1}{2\pi\ell^4_{\rm s}}\int_X \Re(n_a\log\zeta^a-\log \zeta_D)Q^{\rm bg}_6
\ee
Hence, the factor $e^{-S_{\rm D3}}$ appearing in the associated non-perturbative F-terms  is proportional to
\be\label{E3cont}
\cala^{\rm bg}\Big[\prod_{I}\zeta_D(Z_I)\Big]e^{-2\pi n_a\rho^a}
\ee
where we have introduced 
\be
\cala^{\rm bg}=\exp\Big[\frac{1}{\ell^4_{\rm s}}\int_X\big(\log \zeta_D-n_a\log\zeta^a\big)Q^{\rm bg}_6\Big]
\ee
We have added an overall phase in order to complexify  $\cala^{\rm bg}$ in a way naturally  compatible with the bulk complex structure. 

The combination (\ref{E3cont}) is well defined on the total moduli space $\calm_{\rm tot}$, since  $\zeta_D(z)$ is a section of  $\calo_X(D)\simeq \calo_X(n_aD^a)$, while from (\ref{rhotran}) it follows that $e^{-2\pi n_a\rho^a}$ transforms as a section of $\calo_X(-n_aD^a)\simeq \calo_X(-D)$. 

In particular, the presence of the factor $\prod_{I}\zeta_D(Z_I)$, which encodes the dependence on the D3-brane moduli, is in agreement with what was anticipated in \cite{Ganor:1996pe} from an argument based on monodromy and  holomorphy. Hence our procedure  gives,  as particular by-product, an explicit  derivation of such prefactor which is in the same spirit, although more direct and  general,  as the derivation presented in \cite{Baumann:2006th} for certain conifold backgrounds, see also \cite{Dymarsky:2010mf}. 

On the other hand, the factor $\cala^{\rm bg}$ contains the effect of the back-reaction of the other background sources,  see (\ref{Qbg}).
It does not depend on the moduli we consider in the present paper. However, as $\prod_{I}\zeta_D(Z_I)$, it may become relevant in  more general flux-compactifications that include more dynamical fields in the low-energy effective theory.\footnote{For instance, our results automatically incorporate the mechanism proposed in \cite{Marchesano:2009rz} to generate non-perturbatively induced Yukawa couplings.}

%%%%%%%%%%%%%%%%%%%%%%%%%%%%%%%%%%%%%%%%%%%%%%%%%%%%%%%%
%%%%%%%%%%%%%%%%%%%%%%%%%%%%%%%%%%%%%%%%%%%%%%%%%%%%%%%%

\subsection{The $h^{1,1}=1$ case}
\label{sec:single}

Let us restrict to the particular case in which there are no constrained K\"ahler moduli $v_a$, i.e.\ $h^{1,1}(X)=1$, so that  $H^{1,1}(X)$ is generated by the K\"ahler form $J_0$. 
It is then convenient to take it integrally quantised and choose as single integral two-form $\omega\equiv J_0$. This corresponds to setting the single constrained K\"ahler modulus $v\equiv 1$. 
Notice that in this case $h$ is just a constant. Hence it can be omitted from   (\ref{rhoinv}), which then reduces  to 
\be\label{simplrho}
\Re\rho=3{\rm v}_0\hat a+\frac1{2}\sum_I k(Z_I,\bar Z_I)
\ee
Here we have identified $\kappa(z,\bar z)$ with the K\"ahler potential $k(z,\bar z)$ associated with $J_0$, defined by
\be\label{k0pot}
\ii\del\delbar k=J_0
\ee
Inverting (\ref{simplrho}) to write $\hat a=\hat a(\Re\rho,Z,\bar Z)$,  the K\"ahler potential (\ref{complimpl}) takes the form
\be\label{univmodulus}
K=-3\log\Big[\Re\rho-\frac12\sum_I k(Z_I,\bar Z_I)\Big]+ c_0+3\log 3
\ee
This is in  agreement with the K\"ahler potential first proposed in \cite{DeWolfe:2002nn} and further discussed in \cite{Giddings:2005ff,Baumann:2006th,Chen:2009zi}. Notice that we could have started from (\ref{simpk}) and used in  (\ref{simplrho}) the more general universal modulus  $a$  instead of $\hat a=a+\frac{{\rm v}_0^{\rm w}}{{\rm v}_0}$. This shift in the definition of $\Re\rho$ would then modify the functional dependence of $K$ as follows
\be
K=-3\log\Big[\Re\rho-\frac12\sum_I k(Z_I,\bar Z_I)+3{\rm v}^{\rm w}_0\Big]+ c_0+3\log 3
\ee
with ${\rm v}^{\rm w}_0$ as in (\ref{wgauge}), consistently with the results of \cite{Frey:2008xw,Chen:2009zi}.

%%%%%%%%%%%%%%%%%%%%%%%%%%%%%%%%%%%%%%%%%%%%%%%%%%%%%%

\section{Warped kinetic terms and no-scale condition}
\label{sec:kinetic}

In this section we derive the kinetic terms of the chiral fields $\rho^a$ and $Z_I^i$ defined by the K\"ahler potential (\ref{complimpl}). Let us again collectively denote these fields by $\varphi^A$.  Then, according to (\ref{Ekin}) we have to compute the matrix $K_{A\bar B}\equiv \del_A\del_{\bar B}K$.  Since $K$ is only implicitly defined, one needs   to compute the derivatives of $a$ and $v_{a}$ with respect to $\rho^a$ and $Z^i_I$ by using (\ref{rhoinv}). One of the main difficulties comes from the dependence of  $\kappa^a(z,\bar z;v)$ and $h(v)$ on the constrained K\"ahler moduli $v_a$. Nevertheless,  one can actually compute these derivatives. This is discussed in detail in appendix \ref{app:kaehler} and the final results are in (\ref{firstder}) and (\ref{derZ}). Here we just mention that a key ingredient in the derivation is a non-trivial interplay, physically due to the  tadpole condition (\ref{tadpolecond}), between the contributions of $\kappa^a(Z_I,\bar Z_I;v)$ and   $h(v)$.  

As a preliminary step, we  introduce the matrix
\be\label{wM}
M^{ab}_{\rm w}\equiv \int_X e^{-4A}J_0\wedge \omega^a\wedge \omega^b
\ee
Notice that $M^{ab}_{\rm w}$ cannot be written in terms of topological intersection numbers, because of the warping $e^{-4A}$.
Let us also define the warped metric
\be\label{wG}
\calg^{\rm w}_{a b}\equiv\frac{1}{4{\rm v}_0\hat a}\Big[\frac{1}{2{\rm v}_0\hat a}v_av_b-(M^{-1}_{\rm w})_{ab}\Big]
\ee
and
\be
\cala^{aI}_i(Z_I,\bar Z_I;v)\equiv \frac{\del\kappa^a(Z_I,\bar Z_I;v)}{\del Z^{i}_I}\quad~~~~~ \bar\cala^{aI}_{\bar\imath}(Z_I,\bar Z_I;v)\equiv \frac{\del\kappa^a(Z_I,\bar Z_I;v)}{\del \bar Z^{\bar\imath}_I}
\ee
Notice that under the transformations (\ref{genkt}), we have
\be\label{kappatran}
\begin{aligned}
\cala^{aI}_i(Z_I,\bar Z_I;v)&\rightarrow \cala^{aI}_i(Z_I,\bar Z_I;v)+\frac{\del \chi^a(Z_I)}{\del Z^i_I}\\
\bar\cala^{aI}_{\bar\imath}(Z_I,\bar Z_I;v)&\rightarrow \bar\cala^{aI}_{\bar\imath}(Z_I,\bar Z_I;v)+\frac{\del \bar \chi^a(\bar Z_I)}{\del \bar Z^{\bar\imath}_I}
\end{aligned}
\ee

Now, by using (\ref{firstder}) and (\ref{derZ}) one can quite straightforwardly compute the  K\"ahler metric $K_{A\bar B}$ associated with the K\"ahler potential (\ref{complimpl}).
The result is 
\be\label{Kmetric}
K_{A\bar B}\equiv \left(\begin{array}{ccc}  \frac{\del K}{\del \rho^a\del\bar\rho^{ b}}  &  &  \frac{\del K}{\del \rho^a\del \bar Z^{\bar\jmath}_{J}} \\
&&\\
\frac{\del K}{\del Z^{i}_{I}\del \bar\rho^{b}} &  & \frac{\del K}{\del  Z^{i}_{I}\del \bar Z^{\bar\jmath}_{J}}
\end{array}
\right)=
\left(\begin{array}{ccc}  \calg^{\rm w}_{a b}  & &-\calg^{\rm w}_{a c}\bar\cala^{cJ}_{\bar\jmath}  \\
&&\\
-\cala^{cI}_{i}\calg^{\rm w}_{c b}   &  & \calg^{\rm w}_{cd}\cala^{cI}_{i}\bar\cala^{dJ}_{\bar\jmath}  +\frac{1}{2{\rm v}_0\hat a}\delta^{IJ}g_{0,i\bar\jmath}(Z_I,\bar Z_I)
\end{array}
\right)
\ee
where the $g_{0,i\bar\jmath}(z,\bar z)$ is the K\"ahler metric of the internal space $X$.

By using (\ref{Kmetric}) we can write the kinetic terms (\ref{Ekin}) in the form
\be\label{kinlag}
\call_{\rm kin}=-M^2_{\rm P}\calg^{\rm w}_{ab}\nabla_\mu\rho^a \nabla^\mu\bar\rho^b
-\frac{1}{2{\rm v}_0\hat a}M^2_{\rm P}\sum_Ig_{0,i\bar\jmath}(Z_I,\bar Z_I)\del_\mu Z^i_I\del^\mu \bar Z^{\bar\jmath}_I
\ee
where we have introduced the covariant derivatives
\be
\begin{aligned}
\nabla_\mu\rho^a&\equiv \del_\mu\rho^a-\cala^{aI}_i\del_\mu Z^i_I
\end{aligned}
\ee
and $\nabla_\mu\bar\rho^a\equiv (\nabla_\mu\rho^a)^*$, 
which are invariant under the transformations (\ref{rhotran}),  taking (\ref{kappatran}) into account. 

Notice that the second term on the r.h.s\ of (\ref{kinlag})
is  easily reproduced by probe D3-branes. Indeed, focusing on a single D3-brane, from (\ref{partk}) and (\ref{scgauge}) it follows that $M^2_{\rm P}=4\pi {\rm v_0}\hat a|Y|^2$ so that the second term appearing in the r.h.s\ of (\ref{kinlag}) can be written as
\be
-2\pi |Y|^2g_{0,i\bar\jmath}(Z,\bar Z;v)\del_\mu Z^i\del^\mu \bar Z^{\bar\jmath}
\ee
which indeed coincides with the  kinetic term obtained by expanding the DBI action of a probe D3-brane 
 on the ten-dimensional metric (\ref{rescmetric}). This provides a non-trivial  check  of our result.

Now,  from  (\ref{firstder}) and (\ref{derZ}) one can easily compute  the first derivatives of the K\"ahler potential:
\be\label{Kfirstder}
K_A=\left(\begin{array}{ccc} \frac{\del K}{\del\rho^a} &,& \frac{\del K}{\del Z^{i}_I}\end{array}\right)=\left(\begin{array}{ccc}- \frac{v_a}{2v_0\hat a} &,& \frac1{2{\rm v}_0\hat a}v_a\cala^{aI}_i\end{array}\right)
\ee 
and $K_{\bar A}=(K_A)^*$.
Furthermore,  the inverse of the K\"ahler metric (\ref{Kmetric}) takes the form:
 \be\label{invK}
 K^{ A \bar B}=\left(\begin{array}{ccl}  \calg_{\rm w}^{a b}+2{\rm v}_0\hat a\, \sum_Lg_0^{ l \bar m}(Z_L,\bar Z_L)\cala^{aL}_{ l}\bar\cala^{bL}_{ \bar m}  & &2{\rm v}_0\hat a\,\cala^{a L}_{l}g^{ l \bar\jmath}_0(Z_J,\bar Z_J)\delta_{LJ}  \\
&&\\
2{\rm v}_0\hat a\, \delta_{IM} g^{i\bar m}_0(Z_I,\bar Z_I)\bar\cala^{b M}_{\bar m}  &  & 2{\rm v}_0\hat a\,\delta_{IJ}g_0^{i \bar\jmath}(Z_I,\bar Z_I)
\end{array}
\right)
 \ee 
 where $g_0^{\bar\imath j}$ is the inverse of  $g_{0,i\bar\jmath}$ and
 \be
 \calg_{\rm w}^{a b}\equiv\Big(M^{ad}_{\rm w}v_dv_c-4{\rm v}_0\hat a\,\delta^a_c\Big)M^{cb}_{\rm w}
 \ee 
 is  the inverse of $\calg^{\rm w}_{a b}$. 
 Hence, from (\ref{Kfirstder}) and (\ref{invK}) one can straightforwardly check that
 \be\label{noscale}
 K^{A \bar B}K_AK_{\bar B}=3
 \ee
  That is, our warped K\"ahler potential is of {\em no-scale} type \cite{Cremmer:1983bf,Ellis:1983sf}.  
  
In fact (\ref{noscale}) had to be expected. Indeed supersymmetry can be  broken, preserving the ten-dimensional equations of motion, if the classical superpotential (\ref{GVW}) has a 
non-vanishing expectation value $W_0\neq 0$. In particular,  these non-supersymmetric vacua  still have a Minkowski external space.
 The no-scale condition (\ref{noscale}) makes such ten-dimensional supersymmetry breaking mechanism naturally consistent with the effective four-dimensional supergravity viewpoint.   
 Hence, (\ref{noscale}) provides a further consistency check for  our results.

So far we have used only chiral multiplets in the low-energy effective theory. 
On the other hand, one could dualise  the chiral fields $\rho^a$ to linear multiplets. 
Let us denote the scalar components of the dual linear multiplets by $l_a$.
They can be easily computed within the superconformal formalism, before imposing the gauge fixing condition (\ref{scgauge}) on the conformal compensator, and are given by \cite{Ferrara:1983dh,Cecotti:1987nw}  
\be
l_a=\frac{3}{4\pi}\frac{\del\caln}{\del \Re \rho^a}
\ee     
From  (\ref{Nsplit}) it follows  that $\caln=4\pi|Y|^2 {\rm v}_0\hat a$. Then, by using (\ref{dera}) we obtain
\be
l_a= |Y|^2 v_a
\ee
That is, the scalars $l_a$ are just the components appearing in the expansion $J=\ell^2_{\rm s}\,l_a\,\omega^a$ of the non-normalised K\"ahler form $J\equiv \ell^2_{\rm s}|Y|^2J_0$ introduced in (\ref{defJ}). This shows that the  scalars $l_a$ have a more transparent geometrical interpretation, compared to the chiral fields $\rho^a$. The complexity of (\ref{rhoinv}) could  then be regarded as the result of performing the inverse duality from linear to chiral  multiplets. See \cite{Grimm:2004uq,Grimm:2010ks} for a detailed discussion in the unwarped case and \cite{eff2} for similar remarks  in the case of generalised compactifications.

Finally, let us give a closer look at the warped matrix $M^{ab}_{\rm w}$ defined in (\ref{wM}), which encodes the non-trivial warping contribution to $\calg^{\rm w}_{ab}$. In particular, by splitting $e^{-4A}$ as in (\ref{partsplit}), $M^{ab}_{\rm w}$  can be rewritten as
\be\label{splitMw}
M^{ab}_{\rm w}= \hat a\,\cali^{abc}v_{c}+\int_X e^{-4\hat A_0}J_0\wedge \omega^a\wedge \omega^b
\ee
The second term on the r.h.s.\ of this equation is what makes $\calg^{\rm w}_{ab}$ quantitatively different with respect to the corresponding metric in the constant-warping approximation \cite{Grimm:2004uq,Grimm:2010ks}. 

Even though we generically expect the second term on the r.h.s.\ of (\ref{splitMw}) to be non-trivial, there are some particular cases in which it actually vanishes.
Indeed, recall  that   $e^{-4\hat A_0}$ satisfies  the normalisation condition (\ref{norm0}). Hence we see that  the second term on the r.h.s.\ of (\ref{splitMw})
vanishes  if 
\be\label{nowcond}
J_0\wedge \omega^a\wedge \omega^b=C^{ab}\, J_0\wedge J_0\wedge J_0
\ee
for some constants $C^{ab}$. The condition (\ref{nowcond}) is equivalent to requiring that the product  of any harmonic (1,1)-form and any harmonic (2,2)-form  gives a harmonic 6-form.
This is for instance the case for the simple class of $N=1$  compactifications on  $X=T^6/\mathbb{Z}_2$ discussed in appendix \ref{app:tori}.
More generically, such property is  guaranteed if $X$ is geometrically formal \cite{formal} that is, if the product of any pair of harmonic forms gives a harmonic form.
A similar conclusion was reached in \cite{Frey:2013bha} by studying 
the kinetic terms of the $C_4$ axionic moduli on warped Calabi-Yau three-folds.

%%%%%%%%%%%%%%%%%%%%%%%%%%%%%%%%%%%%%%%%%%%%%%%%%%%%%%
%%%%%%%%%%%%%%%%%%%%%%%%%%%%%%%%%%%%%%%%%%%%%%%%%%%%%%

\section{Large moduli limit}
\label{sec:large}

It can be sensible  to consider a regime in which  $\Re\rho^a$ are all very large,  $\Re\rho^a\gg 1$. This is indeed the regime in which the supergravity description of string theory is expected to be fully legitimate. Hence, it is natural to look for a perturbative solution of the inverted relations $\hat a=\hat a(\Re\rho,Z,\bar Z)$ and $v_a=v_a(\Re \rho,Z,\bar Z)$ in powers of $(\Re\rho)^{-1}$.   

First expand $\hat a$ and $v_a$ as follows
\be
\label{expk}
\hat a= \hat a^{\uu{-1}}+\hat a^{\uu{0}}+\calo\Big(\frac{1}{\Re\rho}\Big)\quad~~~~~~~~~~
v_a=v_a^{\uu{0}}+v_a^{\uu{1}}+\calo\Big(\frac{1}{(\Re\rho)^2}\Big)
\ee
Furthermore, by expanding the constraint (\ref{constr})  up to first order, one gets
\be
\label{const1}
v^{\uu{0}}_av^{\uu{0}}_b v^{\uu{0}}_c\,\cali^{abc}=6{\rm v}_0\quad~~~~~~~~~
v^{\uu{0}}_av^{\uu{0}}_b v^{\uu{1}}_c\,\cali^{abc}=0
\ee
Hence the leading order contribution  $\Re\rho^a=\frac12\hat a^{\uu{-1}}\cali^{abc}v_b^{\uu{0}} v^{(0)}_c$ to (\ref{rhoinv})
implicitly gives $\hat a^{\uu{-1}}$ and $v_a^{\uu{0}}$ in terms of $\Re\rho^a$.
On the other hand, the next-to-leading order of (\ref{rhoinv}) gives $a^{\uu{0}}$ in terms of $v_a^{\uu{0}}$:\footnote{One also obtains $v^{\uu{1}}_a=\frac{1}{\hat a^{\uu{-1}}}\Big[\frac1{6{\rm v}_0}v^{\uu{0}}_av^{\uu{0}}_b-(M^{-1}_{\uu{0}})_{ab}\Big]\Big[h^b(v^{\uu{0}})+\frac{1}{2}\sum_I\kappa^b(Z_I,\bar Z_I;v^{\uu{0}})\Big]$.}
\be
\hat a^{\uu{0}}=- \frac{1}{3{\rm v}_0}h(v^{\uu{0}})-\frac{1}{6{\rm v}_0}\sum_I k(Z_I,\bar Z_I;v^{\uu{0}})
\ee
where $k\equiv v_a \kappa^a$ is nothing but the K\"ahler potential associated with $J_0$, as defined in (\ref{k0pot}), and we have introduced the quantities $M_{\uu{0}}^{ab}\equiv\cali^{abc}v_c^{\uu{0}}$ and $h(v^{\uu{0}})\equiv v_a^{\uu{0}} h^a({v^{\uu{0}}})$.

We can now consider $\hat a^{\uu{-1}}$ and $v_a^{\uu{0}}$ as the independent geometric moduli and suppress the superscripts for notational simplicity. Hence, by neglecting terms of order $(\Re\rho)^{-1}$ in the logarithm, the K\"ahler potential   (\ref{complimpl}) can be approximated by 
\be
\label{firstkahler1}
K\simeq -3\log \Big[{\rm v}_0\hat a -\frac1{3} h( v)-\frac1{6}\sum_I k(Z_I,\bar Z_I;v)\Big]+c_0
\ee
where
\be
h(v)\equiv \frac1{2\pi\ell^4_{\rm s}}\int_X\Big[\pi k(z,\bar z;v)-v_a\Re\log\zeta^a(z) \Big]Q^{\rm bg}_6
\ee

Alternatively, by using unconstrained K\"ahler parameters $\hat v_a=v_a\sqrt{\hat a}$ as in  (\ref{vunconst}), associated with  $\hat J=\hat v_a\omega^a$, the K\"ahler potential (\ref{firstkahler1}) can be written as
\be
\label{firstkahler2}
K\simeq -2\log \Big[\frac{1}{3!}\cali^{abc}\hat v_a\hat v_b\hat v_c-\frac1{2} h(\hat v) -\frac1{4}\sum_I k(Z_I,\bar Z_I;\hat v)\Big]-\log{\rm v}_0+c_0
\ee

In (\ref{firstkahler1}) and (\ref{firstkahler2})  the geometric moduli $\hat a$ and $v_a$, or $\hat v_a$,  must be regarded as functions  of $\Re\rho^a$ alone, implicitly defined by
\be\label{firstinv}
\Re\rho^a=\frac12\hat a\,\cali^{abc}v_b v_c\equiv \frac{1}{2}\cali^{abc}\hat v_b\hat v_c
\ee

In (\ref{firstkahler1}), or (\ref{firstkahler2}), the effect of D3-branes and fluxes appears directly in the K\"ahler potential, while the relation (\ref{firstinv}) between geometric moduli $\hat v_a$ and holomorphic moduli $\rho^a$ is formally the same as for constant warping compactifications \cite{Grimm:2004uq,Grimm:2010ks}.  
The modification due to D3-branes is of the same form proposed in \cite{DeWolfe:2002nn}.
On the other hand, the effect of the fluxes and other D3-charge sources is encoded in $h(\hat v)$. Notice that if we rescale $\hat v_a\rightarrow \lambda \hat v_a$,
then $k(z,\bar z;\hat v)$ and  $h(\hat v)$ scale like homogeneous functions of degree one. Hence, by expanding the logarithm, we see that the overall correction to 
the K\"ahler potential scales with $\lambda^{-2}$. In other words, the first order correction  scales like the $V^{-\frac{2}{3}}$, where  $V\equiv \frac{1}{3!}\cali^{abc}\hat v_a\hat v_b\hat v_c$ can be identified with the internal volume of the corresponding 
constant warping compactification.  This scaling behaviour  was also argued in \cite{Frey:2013bha} by studying  the dimensionally reduced theory for  axionic moduli.

%%%%%%%%%%%%%%%%%%%%%%%%%%%%%%%%%%%%%%%%%%%%%%
%%%%%%%%%%%%%%%%%%%%%%%%%%%%%%%%%%%%%%%%%%%%%%%

\section{Including axions}
\label{sec:axions}

In the previous sections we assumed the absence of chiral fields corresponding to purely axionic moduli. In fact, they can be included by using the same strategy followed so far. 
The possible axionic chiral fields are more easily identified in the dual elliptically fibered Calabi-Yau  four-fold $\hat X$, on which they are counted by $b^{3}(\hat X)-b^{3}(X)$, see \cite{Grimm:2010ks} for a detailed discussion. In such description we should use probe Euclidean M5-branes wrapping vertical divisors and, unfortunately,  the M5-brane effective theory on non-trivial supergravity backgrounds is not very manageable \cite{Pasti:1997gx,Bandos:1997ui}. On the other hand, in  the type IIB framework the axionic chiral fields are less easily identified, but one can use the Euclidean D3-brane action, which is under better control.

To avoid such technical complications and outline the general strategy to include axionic moduli, we will restrict to F-theory backgrounds admitting a weakly coupled description,
with D7-branes and O7-planes and an almost constant axio-dilaton $\tau$. 
In this limit one can work in the Calabi-Yau double cover three-fold $\tilde X$ and distinguish between closed and open string axionic chiral fields. 

\subsection{Closed string axionic chiral fields}

Let us first focus on the closed string moduli. Introduce a set of harmonic  (1,1)-forms $\chi_\alpha$  that are odd under the orientifold involution
and  form a basis for   $H^{1,1}_-(\tilde X)$.  We can then fix a certain reference $B_2^{(0)}$ and $C_2^{(0)}$ satisfying the flux equations, and write
$B_2=B_2^{\uu{0}}+\Delta B_2$  and $C_2=C_2^{\uu{0}}+\Delta C_2$, with $\Delta B_2$ and $\Delta C_2$ being closed harmonic $(1,1)$-forms.   We can then expand $\Delta B_2$ and $\Delta C_2$ as follows
\be
\Delta  B_2=\ell^2_{\rm s}\,
b^\alpha\chi_\alpha\quad~~~~~~~~ \Delta  C_2=\ell^2_{\rm s}\,c^\alpha\chi_\alpha
\ee
The moduli $b^\alpha$ and $c^\alpha$ are naturally combined into the following set of  $h^{1,1}_-(\tilde X)$ chiral fields:
\be
\beta^\alpha=c^\alpha-\tau b^\alpha
\ee 
In order to include them in the above discussion, we need to take into account the second term appearing in the on-shell supersymmetric DBI-action (\ref{susyDBI}). 
By considering a Euclidean D3-brane wrapping an even divisor $D\subset \tilde X$ (that is, such that $\sigma(D)=D$), the second term on the r.h.s.\ of (\ref{susyDBI})
can be written as
\be\label{FFexp}
-\frac{\pi}{2\ell^4_{\rm s}}\int_{D}e^{-\phi}\calf_{\uu{0}}\wedge \calf_{\uu{0}}-\Big(\frac{\pi}{\ell^2_{\rm s}}\int_{D}\calf_{\uu{0}}\wedge \chi_\alpha\Big)\Im\beta^\alpha
+\frac{\pi}{\Im\tau}\calm_{\alpha\beta}\,\Im\beta^\alpha\,\Im\beta^\beta
\ee
where $\calm_{\alpha\beta}\equiv -\frac12\int_{D} \chi_\alpha\wedge \chi_\beta$  and we have included an overall factor $\frac12$
to take into account that we are working in the double cover $\tilde X$. 

Taking (\ref{FFexp}) into account,  the definition of $I^a$ given in (\ref{Imoduli}) (written in the covering space) should be modified into
\be\label{Imod}
I^a=\frac{1}{4}\int_De^{-4A}J_0\wedge J_0-\Big(\frac{1}{2\ell^2_{\rm s}}\int_{D}\calf_{\uu{0}}\wedge \chi_\alpha\Big)\Im\beta^\alpha
+\frac{1}{2\Im\tau}\calm_{\alpha\beta}\,\Im\beta^\alpha\,\Im\beta^\beta
\ee
  where we have omitted the contribution of the first term in (\ref{FFexp}), since  moduli independent. The second term in (\ref{Imod}) can be written as  $({\rm hol}+\overline{\rm hol})$. 
Hence, according to the general prescription (\ref{Imoduli2}), it can be ignored as well. Repeating the discussion which lead to (\ref{rhoinv}), we then arrive at the identification
\be\label{crhoinv}
\begin{aligned}
\Re\rho^a&=\hat a\,\calv^a(v)+h^a(v)+\frac1{2}\sum_I \kappa^a(Z_I,\bar Z_I;v)+\frac{1}{2\Im\tau}\calm^{a}_{\alpha\beta}\,\Im\beta^\alpha\,\Im\beta^\beta\\
\end{aligned}
\ee
where 
\be\label{doubleint}
\calm^{a}_{\alpha\beta}\equiv -\frac12\int_{\tilde X}\omega^a\wedge \chi_\alpha\wedge \chi_\beta
\ee

As for the K\"ahler potential, it still takes the form (\ref{complimpl}), up to allowing $\hat a$ to depend also on $\Im\beta^a$.
So (\ref{crhoinv}) is the only modification we need to consider. In particular, the axionic contribution to (\ref{crhoinv})
 is identical to the corresponding contribution in the constant warping approximation \cite{Grimm:2004uq}.

\subsection{Wilson lines}

Finally, we  incorporate the open string axions as well.  Let us focus on a D7-brane wrapping an orientifold invariant divisor $\Sigma=\sigma(\Sigma)$. 
Then, the (orientifold odd) gauge field $A_{\rm D7}$ supported on $\Sigma$  can be split into  $A^{\uu{0}}_{\rm D7}+\Delta A_{\rm D7}$, where $A^{\uu{0}}_{\rm D7}$ is some fixed background gauge field and $\Delta A_{\rm D7}$ is an arbitrary harmonic one-form, which can be regarded as a Wilson line.  
The corresponding chiral fields $a^\cali$ can be obtained by expanding the $(0,1)$ component 
 $\Delta A^{0,1}_{\rm D7}$ in a basis of harmonic (0,1)-forms $\eta_\cali\in H^{0,1}(\Sigma)$:
 \be\label{wilsonexp}
 \Delta A^{0,1}_{\rm D7}=2\pi a^\cali\,\eta_\cali
 \ee
 
 Now, the gauge field $A_{\rm D7}$ couples to a  Euclidean D3-brane wrapping a divisor $D$ 
 through its coupling with the chiral scalar $\beta$ supported on the intersection curve $\calc=\Sigma\cap D$ \cite{Martucci:2014ema}:
 \be\label{chiralboson}
\frac\ii{8\pi}\int_{\calc} \Big[\del\beta\wedge\delbar\beta-2\del\beta\wedge (A_{\rm D3}^{0,1}-A_{\rm D7}^{0,1})+A_{\rm D3}^{1,0}\wedge A_{\rm D3}^{0,1}-2A^{1,0}_{\rm D3}\wedge A^{0,1}_{\rm D7}
+A_{\rm D7}^{1,0}\wedge A_{\rm D7}^{0,1}\Big]
 \ee
This two-dimensional action contributes to $S_{\rm D3}$.  In order to identify the modification of (\ref{crhoinv}) due to Wilson lines, we use the same strategy as above and we require  $S_{\rm D3}$ to depend homomorphically on the background chiral fields. The terms that  do not depend on $A_{\rm D7}$ can be ignored. Furthermore, the terms that are linear on $A_{\rm D7}^{0,1}$ can be ignored as well,  since they depend holomorphically on 
the Wilson lines $a^I$. On the other hand,  the real quadratic term 
\be\label{wilson}
\frac\ii{8\pi}\int_{\calc} A_{\rm D7}^{1,0}\wedge A_{\rm D7}^{0,1}
\ee
 must be taken into account.  More specifically, by using the expansion (\ref{wilsonexp}),   the definition of $\Re\rho^a$ given in (\ref{crhoinv}) must be further  modified into
 \be\label{crhoinv2}
\begin{aligned}
\Re\rho^a&=\hat a\,\calv^a(v)+h^a(v)+\frac1{2}\sum_I \kappa^a(Z_I,\bar Z_I;v)+\frac{1}{2\Im\tau}\calm^{a}_{\alpha\beta}\,\Im\beta^\alpha\,\Im\beta^\beta+\frac{1}{2}\calc^{a}_{\cali\calj}a^{\cali}\bar a^{\calj}
\end{aligned}
\ee
where
\be
\calc^{a}_{\cali\calj}=-\frac{\ii}{2}\int_{\calc^a}\eta_\cali\wedge \bar\eta_{\calj}
\ee
with $\calc^a\equiv D^a\cap\Sigma$.
The formula (\ref{crhoinv2}) can be obviously extended to include the contribution of additional D7-branes. Notice that the contribution of the Wilson lines to (\ref{crhoinv2}) has the same form of the contribution identified  in  \cite{Jockers:2004yj} by using probe D7-branes on  unwarped compactifications.

\subsection{Complete K\"ahler potential for $h^{1,1}_+=1$}

The new relations(\ref{crhoinv2}) should be partially inverted  in order to find $\hat a$ as function of the chiral fields $\rho^a$, $Z^i_I$, $\beta^\alpha$ and $a^\cali$, and to properly interpret   the K\"ahler potential  as a function of the chiral coordinates.
The case $h^{1,1}_+=1$ is the simplest one in which one can explicitly invert (\ref{crhoinv2}). Indeed, proceeding as in section \ref{sec:single} one arrives at the K\"ahler potential
\be\label{univaxion}
K=-3\log\Big[\Re\rho-\frac1{2}\sum_I k(Z_I,\bar Z_I)-\frac{1}{2\Im\tau}\calm_{\alpha\beta}\,\Im\beta^\alpha\,\Im\beta^\beta-\frac{1}{2}\calc_{\cali\bar\calj}a^{\cali}\bar a^{\bar\calj}\Big]+c_0+3\log 3
\ee
If we had used in (\ref{crhoinv2}) the more general universal modulus $a$ instead of $\hat a$, we would have obtained the K\"ahler potential
\be
K=-3\log\Big[\Re\rho-\frac1{2}\sum_I k(Z_I,\bar Z_I)-\frac{1}{2\Im\tau}\calm_{\alpha\beta}\,\Im\beta^\alpha\,\Im\beta^\beta-\frac{1}{2}\calc_{\cali\bar\calj}a^{\cali}\bar a^{\bar\calj}+3{\rm v}^{\rm w}_0\Big]+c_0+3\log 3
\ee
with ${\rm v}^{\rm w}_0$ defined in (\ref{wgauge}).
Ignoring the contribution of  mobile D3-branes and D7-brane Wilson lines, this matches the K\"ahler potential obtained in \cite{eff2}, using a strategy similar to the one followed in the present paper, which was then confirmed by a direct dimensional reduction in \cite{Frey:2013bha}.

%%%%%%%%%%%%%%%%%%%%%%%%%%%%%%%%%%%%%%%%%%%%%%%%%%%%%

\subsection{K\"ahler metric and no-scale structure}

In order to compute the K\"ahler metric one must use the derivatives of the geometric moduli discussed in appendix \ref{app:axions}.
The final result is
\be\label{kinlagaxion}
\begin{aligned}
\call_{\rm kin}=&-M^2_{\rm P}\,\calg^{\rm w}_{ab}\nabla_\mu\rho^a \nabla^\mu\bar\rho^b
-\frac{1}{2{\rm v}_0\hat a}M^2_{\rm P}\sum_Ig_{0,i\bar\jmath}(Z_I,\bar Z_I)\del_\mu Z^i_I\del^\mu \bar Z^{\bar\jmath}_I+\\
&\quad~~~~~~~-\frac{1}{4{\rm v}_0\hat a \Im\tau}M^2_{\rm P}\,v_a\calm^a_{\alpha\beta}\del_\mu\beta^\alpha\del^\mu\bar\beta^\beta-\frac{1}{2{\rm v}_0\hat a}M^2_{\rm P}\,v_a\calc^a_{\cali\calj}\del_\mu a^\cali\del^\mu\bar a^\calj
\end{aligned}
\ee
where $\calg^{\rm w}_{ab}$ is as in (\ref{wG})  and the covariant derivatives  are now defined as follows:
\be
\begin{aligned}
\nabla_\mu\rho^a&\equiv \del_\mu\rho^a-\cala^{aI}_i\del_\mu Z^i_I+\frac{\ii}{\Im\tau}\calm^a_{\alpha\beta}\Im\beta^\alpha\del_\mu\beta^\beta    -\calc^a_{\cali\calj}\bar a^\calj\del_\mu a^\cali
\end{aligned}
\ee
and $\nabla_\mu\bar\rho^a\equiv (\nabla_\mu\rho^a)^*$. 

By ignoring D3-branes, in the constant warping approximation the kinetic terms (\ref{kinlagaxion}) are consistent  with the results of \cite{Jockers:2004yj}. In particular, as a direct check of the numerical factors, the last term on the r.h.s.\ of (\ref{kinlagaxion}) can be easily reproduced by dimensionally reducing the DBI action for a probe D7-brane on the metric (\ref{rescmetric}).

Furthermore, one can verify that the no-scale condition (\ref{noscale}) is still satisfied, as required by physical consistency.

Finally, by performing  the large moduli expansion discussed in section \ref{sec:large}, one can  obtain the appropriate modification of (\ref{firstkahler1}) or, equivalenty,  of (\ref{firstkahler2}). For instance, the latter reads
\be
\label{firstkahler3}
\begin{aligned}
K\simeq& -2\log \Big[\frac{1}{3!}\cali^{abc}\hat v_a\hat v_b\hat v_c -\frac1{2} h(\hat v)-\frac1{4}\sum_I k(Z_I,\bar Z_I;\hat v)\\
&\quad~~~~~~~~~~~-\frac{1}{4\Im\tau}\hat v_a\calm^a_{\alpha\beta}\Im\beta^\alpha\Im\beta^\beta-\frac14 \hat v_a\calc^a_{\cali\calj}a^\cali\bar a^\calj\Big]-\log{\rm v}_0+c_0
\end{aligned}
\ee
where  $\hat v_a$ are functions of $\Re\rho^a$ implicitly determined by $\Re\rho^a= \frac{1}{2}\cali^{abc}\hat v_b\hat v_c$.

%%%%%%%%%%%%%%%%%%%%%%%%%%%%%%%%%%%%%%%%%%%%%%%%%%%%%
%%%%%%%%%%%%%%%%%%%%%%%%%%%%%%%%%%%%%%%%%%%%%%%%%%%%%%%

\vspace{2cm}

\centerline{\large\em Acknowledgments}

\vspace{0.2cm}

\noindent I would like to thank S.~Andriolo, S.~Giusto, E.~Plauschinn, R.~Savelli, R.~Valandro and especially  T.~Weigand for useful discussions and comments on the draft.
The author is grateful to the Mainz Institute for Theoretical Physics (MITP) for its hospitality and its partial support during the completion of this work. This work
is supported in part by the Padua University Project CPDA144437.

\vspace{0.5cm}

\begin{appendix}

\section{Derivatives  of geometric moduli}
\label{app:kaehler}

The relations (\ref{rhoinv}), (\ref{crhoinv}) and (\ref{crhoinv2}) implicitly define the geometric moduli $\hat a$ and $v_a$ as functions of the chiral fields $\rho^a$, $Z^i_I$, $\beta^a$ and $a^\cali$. 
Even though in general one cannot make such dependence explicit, it is possible to compute the derivatives of $\hat a$ and $v_a$ with respect to the chiral fields.  

\subsection{Useful preliminary formulas} 

Once we have chosen the integral two-forms $\omega^a$ to be $(1,1)$ and harmonic, we can expand the normalised K\"ahler form as 
\be
J_0=v_a\,\omega^a
\ee
Notice that the $\omega^a$'s, being harmonic, depend on the constrained K\"ahler moduli $v_a$. Now, since any K\"ahler structure deformation should preserve
 (\ref{ricci}), one can easily adapt the discussion for Ricci-flat Calabi-Yau spaces \cite{Candelas:1990pi} to argue  that any infinitesimal K\"ahler structure deformation must take the form
 \be\label{varJ}
 \delta J_0=\delta v_a\,\omega^a
 \ee
By consistency, this implies that
\be\label{constcond}
v_a\,\delta\omega^a=0
\ee
Moreover the normalisation condition (\ref{constr})  implies that
\be\label{varconst}
\cali^{abc}v_av_b\delta v_b=0
\ee
where $\cali^{abc}$ are the intersection numbers defined in (\ref{intn}).

Another useful identity is
\be\label{jja}
\frac12J_0\wedge J_0\wedge \omega^a=\frac{1}{3!}(J_0\lrcorner \omega^a)J_0\wedge J_0\wedge J_0=d^a(v)\d{\rm vol}_{X,0}
\ee
where
\be\label{constca}
d^a(v)\equiv J_0\lrcorner \omega^a\equiv \frac{1}{2{\rm v}_0}\cali^{abc}v_bv_c
\ee
The constraint (\ref{varconst}) implies $\delta v_a d^a=0$
and then 
\be\label{cvolume}
J_0\wedge J_0\wedge \delta J_0=\delta v_a\, J_0\wedge J_0\wedge \omega^a=0
\ee
In other words, the normalisation condition (\ref{constr}) requires $\delta J_0$ to be primitive. 

In the following computations one of the main subtleties comes from the dependence of the potentials $\kappa^a(z,\bar z;v)$, 
defined by (\ref{defkappa}), on the constrained K\"ahler moduli $v_a$.  Under a deformation $\delta v_a$, we can write 
\be\label{varomega}
\delta\omega^a=\ii\del\delbar\delta\kappa^a
\ee
where, since the cohomology class of $\omega^a$ is fixed, the variations $\delta\kappa^a$ are {\em globally} defined functions on $X$.
Now, by  using (\ref{varomega}) and (\ref{varJ}), the constancy of (\ref{constca}) implies the following equation for $\delta\kappa^a$
\be\label{kappalaplace}
\Delta_0\delta\kappa^a=-2(\delta d^a+\delta v_b\,\omega^a\lrcorner\omega^b)
\ee
Consistently, the r.h.s\ is a coexact function since from (\ref{constca}) one can check that 
\be
\delta d^a=\frac{1}{{\rm v}_0}\cali^{abc}v_b\delta v_c=-\frac{1}{{\rm v}_0}\delta v_b\int_X \omega^a\lrcorner\omega^b\d{\rm vol}_{X,0}
\ee
Eq.~(\ref{kappalaplace}) can be integrated, up to an irrelevant constant along $X$, by means of the Green's function $G(y,y')$ introduced in section \ref{sec:universal}:
\be\label{deltakappa}
\delta\kappa^a(y;v)=2\delta v_b\int_{X,y'} G(y;y')\big(J_0\wedge \omega^a\wedge\omega^b\big)(y')
\ee
Notice that from  (\ref{cvolume}) it follows that $v_a\delta\kappa^a=0$, consistently with (\ref{constcond}).

Now, recalling the definition (\ref{defh}) of $h^a(v)$,
observe that 
\be
\delta h^a(v)+\frac1{2}\sum_I \delta\kappa^a(Z_I,\bar Z_I;v)=\frac{1}{2\ell^4_{\rm s}}\int_X \delta\kappa^a Q_6
\ee 
Combining this equation with  (\ref{deltakappa}) and (\ref{warpgreen}), one can  deduce the following crucial formula:\footnote{In deducing the formula (\ref{cformula}) we have neglected the possible dependence  on the K\"ahler moduli of the higher-order curvature corrections appearing in $Q_6$. Taking such dependence into account  would require to work beyond the second derivative order in the effective action.}
\be\label{cformula}
\delta h^a(v)+\frac1{2}\sum_I \delta\kappa^a(Z_I,\bar Z_I;v)=\delta v_b\int_X e^{-4\hat A_0}J_0\wedge \omega^a\wedge\omega^b
\ee

\subsection{Derivatives in absence of axionic chiral fields}

We are now ready to compute the derivatives of the background geometric moduli with respect to the chiral fields.
Take first the derivatives of  (\ref{rhoinv}) with respect to $\Re\rho^b$. By using (\ref{cformula}) one gets
\be\label{derho}
\delta^a_b=\frac{\del\hat a}{\del\Re\rho^b}\calv^a+\cali_{\rm w}^{acd}v_c\frac{\del v_d}{\del \Re\rho^b}
\ee
where we have introduced the warped `intersection numbers'
\be
\cali^{abc}_{\rm w}\equiv \int_Xe^{-4A}\omega^a\wedge \omega^b\wedge \omega^c
\ee
Notice that $\cali^{abc}_{\rm w}$ is not a topological quantity and that by (\ref{cvolume}) we have
\be\label{wvarconst}
\cali_{\rm w}^{abc}v_av_b\delta v_b=0
\ee
Let us also define
\be
M^{ab}_{\rm w}\equiv \int_Xe^{-4A}J_0\wedge\omega^a\wedge \omega^b\equiv \cali_{\rm w}^{abc}v_c
\ee
Then from (\ref{derho}) one can derive
\begin{subequations}\label{firstder}
\begin{align}
\frac{\del\hat a}{\del\Re\rho^a}&=\frac{v_a}{3{\rm v}_0}\label{dera}\\
\frac{\del v_a}{\del \Re\rho^b}&=(M^{-1}_{\rm w})_{ab}-\frac{1}{6{\rm v}_0\hat a}\, v_av_b\label{derv}
\end{align}
\end{subequations}
Indeed, (\ref{dera}) can be obtained by contracting (\ref{derho}) with $v_a$ and using (\ref{wvarconst}).
Then, one can determine (\ref{derv}) by substituting (\ref{dera}) in (\ref{derho}) and by observing that 
\be
M^{ab}_{\rm w}v_b=2\hat a\,\calv^a(v)
\ee
which follows from  (\ref{jja}) and (\ref{norm0}).

Starting back from (\ref{rhoinv}), taking its derivative with respect to $Z^i_I$ and using again (\ref{cformula})  one gets
\be\label{derz}
0=\frac{\del\hat a}{\del Z^i_I}\calv^a+ M^{ab}_{\rm w}\frac{\del v_b}{\del Z^i_I}+\frac{1}{2}\frac{\del \kappa^a(Z_I,\bar Z_I;v)}{\del Z_I^i}
\ee
Then, proceeding as in the previous case, we arrive at
\begin{subequations}\label{derZ}
\begin{align}
\frac{\del \hat a}{\del Z^i_I}&=-\frac{v_a}{6{\rm v}_0}\frac{\del \kappa^a(Z_I,\bar Z_I;v)}{\del Z^i_I}\\
\frac{\del v_a}{\del Z^i_I}&=\frac{1}{2}\Big[\frac{1}{6{\rm v}_0\hat a}\, v_av_b-(M^{-1}_{\rm w})_{ab}\Big]\frac{\del \kappa^b(Z_I,\bar Z_I;v)}{\del Z^i_I}
\end{align}
\end{subequations}

%%%%%%%%%%%%%%%%%%%%%%%%%%%%%%

\subsection{Derivatives in presence of  axionic chiral fields}
\label{app:axions}

The discussion of the previous subsection can be repeated starting from (\ref{crhoinv}) or more generically (\ref{crhoinv2}). Let collectively denote the additional chiral fields $\beta^a$ and $a^\cali$ with $\phi^{\mathbb{A}}$ and rewrite (\ref{crhoinv}) or  (\ref{crhoinv2})  in the general form
\be\label{rhoax}
\Re\rho^a=\hat a\,\calv^a(v)+\frac1{2}\sum_I \kappa^a(Z_I,\bar Z_I;v)+h^a(v)+N^a(\phi,\bar\phi)
\ee
where $N^a(\phi,\bar\phi)$ {\em does not} depend on the chiral fields $\rho^a$ and $Z^i_I$. Proceeding as in the previous subsection one again arrives at (\ref{firstder})
and (\ref{derZ}). In addition, one also gets 
\begin{subequations}\label{derphi}
\begin{align}
\frac{\del \hat a}{\del\phi^{\mathbb{A}}}&=-\frac{1}{3{\rm v}_0}v_aN^a_{\mathbb{A}}\\
\frac{\del v_a}{\del \phi^{\mathbb{A}}}&=\Big[\frac{1}{6{\rm v}_0\hat a}\, v_av_b-(M^{-1}_{\rm w})_{ab}\Big]N^b_{\mathbb{A}}
\end{align}
\end{subequations}
where $N_{\mathbb{A}}\equiv \frac{\del N^a}{\del\phi^{\mathbb{A}}}$.

%%%%%%%%%%%%%%%%%%%%%%%%%%%%%%%%%

%%%%%%%%%%%%%%%%%%%%%%%%%%%%%%%%%%%%%%%%%%%%%%%%%%%%%%

\section{A simple example}
\label{app:tori}

 In this appendix we apply our results to a simple concrete class of $N=1$  compactifications on  $X=T^6/\mathbb{Z}_2$ \cite{Kachru:2002he}, for which the K\"ahler potential can be made fully explicit. The internal six-torus is factorised, $T^6=T^2\times T^2\times T^2$, and the corresponding complex coordinates $(z^1,z^2,z^3)$ are periodically
identified according to the rule $z^i\simeq z^i+n^i+\lambda m^i$ (no sum over $i$), where $n^i,m^i\in\mathbb{Z}$ and $\lambda\equiv e^{\frac{2\pi\ii}{3}}$ defines the complex structure modulus of the three tori. The axio-dilaton takes the value $\tau=e^{\frac{2\pi\ii}{3}}$ too.
The $\mathbb{Z}_2$ orientifold involution is defined by $z^i\rightarrow -z^i$. One can choose a supersymmetric $G_3$ flux of the form 
\be\label{g3T}
G_3\sim \d z^1\wedge \d z^2\wedge \d\bar z^3+\d z^2\wedge \d z^3\wedge \d\bar z^1+\d z^3\wedge \d z^1\wedge \d\bar z^1
\ee
The flux quantisation condition  fixes the number of D3-branes. In \cite{Kachru:2002he} one can find an example with $N_{\rm D3}=10$. 

On $T^6/\mathbb{Z}_2$ one has $h^{1,1}_+=9$, but  the primitivity requirement $J_0\wedge G_3=0$ imposes additional  conditions. Such conditions are not explicitly taken into account in the rest of the present paper but, as this appendix shows, they can be easily incorporated.  

Even though one would generically expect six non-trivial conditions from $J_0\wedge G_3=0$, the simplicity of (\ref{g3T}) implies that only three conditions are effective. 
These select six out of the nine independent integral even $(1,1)$ forms, which can be chosen as follows:
\be
\begin{aligned}
&\omega^1=\frac{\ii}{\Im\lambda}\d z^1\wedge \d \bar z^1\,,\quad~~~ \omega^2=\frac{\ii}{\Im\lambda}\d z^2\wedge \d \bar z^2\,,\quad~~~ \omega^3=\frac{\ii}{\Im\lambda}\d z^3\wedge \d \bar z^3\quad~~~ ,\\
&\omega^{4}=\frac{1}{\Im\lambda}\Im(\d \bar z^2\wedge\d z^{3})\,,\quad \omega^{5}=\frac{1}{\Im\lambda}\Im(\d \bar z^3\wedge\d z^{1})\,,\quad \omega^{6}=\frac{1}{\Im\lambda}
\Im(\d \bar z^1\wedge\d z^{2})
\end{aligned}
\ee
The non-vanishing  intersection numbers $\cali^{abc}=\frac12\int_{T^6}\omega^a\wedge\omega^b\wedge\omega^c$ are given by
\be\label{intnumb}
\cali^{123}=4\ ,\quad \cali^{456}=1\ ,\quad \cali^{144}=\cali^{255}=\cali^{366}=-2
\ee
Then the constraint (\ref{constr}) takes the form
\be\label{T6con}
4v_1v_2v_3 +v_4v_5v_6-v_1v_4^2-v_2v_5^2-v_3v_6^2={\rm v}_0
\ee

Since the K\"ahler form $J_0=v_a\omega^a$ as well as the two-forms $\omega^a$ have constant coefficients, $\omega^a$ are automatically harmonic. In particular
 they do not depend on the constrained K\"ahler moduli and the potentials
\be
\begin{aligned}
&\kappa^1=\frac{z^1\bar z^1}{\Im\lambda}\quad~~~~~~\kappa^2=\frac{z^2\bar z^2}{\Im\lambda}\quad~~~~~~\kappa^3=\frac{z^3\bar z^3}{\Im\lambda}\\
&\kappa^4=\frac{\Re(z^2\bar z^3)}{\Im\lambda}\quad~~~~~~\kappa^5=\frac{\Re(z^3\bar z^1)}{\Im\lambda}\quad~~~~~~\kappa^6=\frac{\Re(z^1\bar z^2)}{\Im\lambda}
\end{aligned}
\ee 
do not depend on the moduli $v_a$ either.  This implies the drastic simplification that the functions $h^a(v)$ defined in (\ref{defh}) 
are constant, so that  they can be ignored in (\ref{rhoinv}).

Hence,  the geometric moduli $\hat a$ and $v_a$ are related to the chiral fields $\rho^a$ and $Z^i_I$ by
\be\label{t6v}
\begin{aligned}
\hat a(4v_2v_3-v_4^2)&=\Re\rho^1-\frac{1}{2\Im\lambda} \sum_I Z^1_I\bar Z^1_I\equiv T^1\\
\hat a(4v_1v_3-v_5^2)&=\Re\rho^2-\frac{1}{2\Im\lambda} \sum_I Z^2_I\bar Z^2_I\equiv T^2\\
\hat a(4v_1v_2-v_6^2)&=\Re\rho^3-\frac{1}{2\Im\lambda} \sum_I Z^3_I\bar Z^3_I\equiv T^3\\
\hat a(v_5v_6-2v_1 v_4)&=\Re\rho^4-\frac{1}{2\Im\lambda} \sum_I \Re(Z^2_I\bar Z^3_I)\equiv T^4\\
\hat a(v_4v_6-2v_2 v_5)&=\Re\rho^5-\frac{1}{2\Im\lambda} \sum_I \Re(Z^3_I\bar Z^1_I)\equiv T^5\\
\hat a(v_4v_5-2v_3 v_6)&=\Re\rho^6-\frac{1}{2\Im\lambda} \sum_I \Re(Z^1_I\bar Z^2_I)\equiv T^6
\end{aligned}
\ee
where we have introduced the set of real functions 
\be
T^a(\Re\rho,Z,\bar Z)\equiv\Re\rho^a-\frac12\sum_I\kappa^a(Z_I,\bar Z_I)
\ee 
By using (\ref{T6con}) one can  partially invert the relations (\ref{t6v}) to find the function $\hat a(\Re\rho, Z,\bar Z)$
and then write the K\"ahler potential (\ref{complimpl}) as follows:
\be\label{T6K}
K=-\log\Big[T^1T^2T^3+2T^4T^5T^6-T^1\big(T^4\big)^2-T^2\big(T^5\big)^2-T^3\big(T^6\big)^2\Big]-\log\frac{\rm v_0}4+c_0
\ee
So for these simple models one can find a fully explicit K\"ahler potential. On the other hand, these models  are   too simple to exhibit non-trivial
warping  effects due to the fluxes.

\end{appendix}

%\maketitle  IS IGNORED %%%%%%%%%%%

%\listoftables       % ONLY IN DRAFT MODE
%\listoffigures      % ONLY IN DRAFT MODE

%\bibliographystyle{abe}
%\bibliography{references}{}

\providecommand{\href}[2]{#2}\begingroup\raggedright\endgroup

%%%%%%%%%%%%%%%%%%%%%%%%%%%%%%%%%%%%%%%%%%%%%%%%%

\end{document}